\documentclass[12pt,preprint,preprintnumbers,amsmath,amssymb,nofootinbib,aps]{revtex4}

\usepackage{graphics,color,array,dcolumn}
\usepackage{hyperref}
\usepackage{amsmath}
\usepackage{amssymb}
\usepackage{amsfonts}

\def\be{\begin{equation}}
\def\ee{\end{equation}}
\def\ben{\begin{displaymath}}
\def\een{\end{displaymath}}
\def\bea{\begin{eqnarray}}
\def\eea{\end{eqnarray}}

\newcommand{\tr}{{\mathrm{tr}}}

\newcommand{\Diff}{\mathit{Diff}}

\def\be{\begin{equation}}
\def\ee{\end{equation}}
\def\bea{\begin{eqnarray}}
\def\eea{\end{eqnarray}}

\def\R{{\mathbb R}}
\def\C{{\mathbb C}}

\def\x2{X}

\def\s2{{\sqrt 2}}

\begin{document}


\title{\vspace*{4mm}{\bf Gravity and Unification: A review\\[3ex]}}

\author{K. Krasnov$^a$}\email{kirill.krasnov@nottingham.ac.uk}
\author{R. Percacci$^b$}\email{percacci@sissa.it}

\affiliation{{\footnotesize $^a\,$School of Mathematical Sciences,
University of Nottingham, Nottingham, NG7 2RD, UK\\
$^b\,$ SISSA, via Bonomea 265, I-34136 Trieste, Italy\\
and INFN, Sezione di Trieste, Italy\\}}
\date{December 2017}

\begin{abstract}
\noindent 
We review various classical unified theories of gravity and other interactions that have appeared in the literature, paying special attention to scenarios in which spacetime remains four-dimensional, 
while an ``internal'' space is enlarged. The starting point for each such unification scenario is a particular formalism for General Relativity.
We thus start by reviewing, besides the usual Einstein-Hilbert and Palatini formulations, the Einstein-Cartan, MacDowell-Mansouri and BF (both non-chiral and chiral) formulations. Each of these introduces some version of ``internal'' bundle and a dynamical variable that ties the internal and tangent bundles. In each of these formulations there is also an independent connection in the ``internal'' bundle. One can then study the effects of ``enlarging the internal space'', which typically leads to a theory of gravity and Yang-Mills fields. We review what has been done in the literature on each of these unification schemes, and compare and contrast their achievements to those of the better developed Kaluza-Klein scenario. 
\end{abstract}
\maketitle

\section{Introduction}

Both General Relativity (GR) and Yang-Mills (YM) theories are geometric. It is thus not surprising that there have been attempts to unify them in the framework of some classical field theory. There was a flurry of activity along these lines in the context of the Kaluza-Klein scenario in the 70's and 80's. This activity was later subsumed by the development of string theory. The developments in the latter during the last two decades have led the majority of the theoretical physics community to abandon the idea of gravity - YM unification at the level of classical theory as too naive. Instead, the currently prevailing view is that both gravity and YM arise naturally in the context of string theory, with gravity being the low energy limit of closed strings, and YM being the low energy limit of open strings. This does imply relations between the two theories, but these relations are very different from the idea that gravity and YM are parts of a single classical theory. Instead, (super)gravity on a certain background
manifold with boundary is equivalent to a certain (super)-YM theory on the boundary \cite{Witten:1998qj}. In a different relation between the two theories, 
gravity can be seen as YM theory squared \cite{Bern:2010ue}. 

At the same time, the ideas of unification at the level of classical theory are almost as old as the subject of GR itself, as we shall review shortly. 
It may well be that, as in many other cases, 
history will eventually make another full circle and these ideas 
will attract attention again. 
The aim of the present review is to collect what is currently known about the subject, and compare and contrast different approaches, in the hope that this will create a useful resource for future developments. 

The history of attempts at unification of gravity with other forces of Nature started shortly after the formulation of the GR itself \cite{weyl,kaluza,klein}.
Over almost four decades, many different routes have been tried by Einstein himself, as well as researchers influenced by his ideas,  see \cite{Goenner:2004se,Goenner:2014mka} for the history of Einstein's attempts and related works. 
The common consensus is that none of these approaches succeeded.

In retrospect, at least one of the reasons for this was that researchers did not have all the ingredients to be unified.\footnote{There was also Einstein's insistence on availability of everywhere regular solutions he wanted to use to describe matter, as well as his reluctance to accept the unavoidability of the quantum theory, see \cite{Goenner:2014mka}.}  Indeed, Einstein was only concerned with the unification of gravity with Maxwell theory.\footnote{Schr\"odinger, on the other hand, tried to include a ``meson'' field into his unified theory, see \cite{Goenner:2014mka}, Section 8.}  Einstein died in 1955. Just a year before his death YM theory was proposed.
YM theory came to prominence with the discovery of the Higgs mechanism in 1965. In 1967 the electroweak unified theory was proposed. Asymptotic freedom was discovered in 1973 and suggested that also the strong interactions can be described by non-Abelian gauge fields. Thus, Einstein did not live long enough to have all the pieces of the unification puzzle.

In spite of this, many of the ideas that were proposed
in those early efforts evolved and survived in some form until the present day. 
For example, it is noteworthy that $SO(3)$ gauge fields were discovered one year before Yang and Mills by Pauli in the context of the Kaluza-Klein reduction of what we would now call a sphere bundle
\cite{Straumann:2000zc}.\footnote{It is interesting to remark that the notions of fibre bundle, connection and parallel transport was already familiar to mathematicians at that time, it had just not penetrated to physics yet. A particularly relevant example is the notion of Ehresmann connection introduced in 1950 \cite{Ehresmann}. This defines a connection as the horizontal subbundle of the tangent bundle of the total space of the fiber bundle. It is this notion of the connection that arises most naturally in the context of Kaluza-Klein dimensional reduction, where the horizontal distribution appears geometrically as the orthogonal complement to the vertical one. } The work of Kaluza and Klein was extended to general non-abelian groups in the 1960's \cite{dewitt,kerner}.
Its revival in the late 1970's and early 1980's was a prelude to
the subsequent development of string theory.
The emergence of gauge theories was also heavily influenced by the
work of Weyl on his unified theory \cite{oraif}.

As we already mentioned above, unification of the non-gravitational forces of Nature with gravity in the form of a classical field theory is not, at least currently, a popular topic. This is in part due to the discovery of new relations between gravity and YM
\cite{Witten:1998qj}, \cite{Bern:2010ue}. 
Further reasons include: (i) the stigma associated to the unification idea by Einstein's unsuccessful attempts; (ii) somewhat disappointing conclusions that resulted from the revival of the Kaluza-Klein theories in the 1980's, see Section \ref{sec:KK}; (iii) expectation that no classical unified theory of this sort can be promoted to a UV complete quantum theory. Nevertheless, papers on this topic do appear from time to time, often to rediscover what was earlier done by other authors. We hope this exposition will make the existing constructions better known and thus save researchers from rediscovering them in the future.

In one way or another, all (classical) approaches to the unification of gravity proceed by taking some structure from pure gravity, enlarging or generalising it in a geometrically natural way and then reinterpreting the added structures in terms of other physically relevant fields.
These unification proposals can then be classified according to their starting point.

Thus, all the unification attempts that are the focus of this review
will be in correspondence with formalisms for GR that lie at their starting points. 
These fall in the following three broad categories: (i) formulations of second-order in derivatives, in which the spacetime metric (or a field that encodes the metric, see below) is the only field appearing in the Lagrangian; (ii) first-order formulation where in addition to the metric also an independent connection field is introduced; (iii) second-order formulations in which the metric variable of the first-order formalism is integrated out (this only works with a non-zero cosmological constant). 

In the first category, where the spacetime metric is the only field to play with, there is not much room for generalisation. One possibility is to stay in four dimensions but remove the assumption of symmetry
of the basic field. This possibility has been studied extensively by Einstein and co-workers, see \cite{Goenner:2014mka}. Another prominent contributor was Schr\"odinger, even though his work involved also an independent connection field. This will be commented upon later on. The idea of combining the metric with an anti-symmetric tensor has been persistently criticised by Pauli as going against the spirit of unification: ``What God separated, the humans must not join``, see \cite{Goenner:2014mka}, page 67. This idea, however, has survived up to the present day. The anti-symmetric tensor that can be put together with the symmetric metric has become known as the $B$-field, or 2-form field. The $B$-field appears naturally in many contexts. For example, it is an important part of Hitchin's generalised geometry \cite{Hitchin:2004ut}, where it becomes unified with the metric. It is also a necessary ingredient of double field theory \cite{Hull:2009mi}. Still, it is clear that a $4\times 4$ tensor does not have enough components to contain all the bosonic fields that are the present in the Standard Model (SM) of particle physics. 
\footnote{Replacing the metric by a frame field (or tetrad, or vierbein) does not change
this conclusion, since the additional degrees of freedom
are pure gauge.}

The only other possibility in the metric context is to allow for a higher number of spacetime dimensions. This leads to the Kaluza-Klein scenario, which is still one of the most popular approaches to unification. Given that the dimension of the internal space is in principle unlimited,\footnote{Unless supersymmetry is assumed; supergravity can exist in at most 11D.} there is as much space here as one wants, and so in principle all the known gauge fields, together with gravity, can be accounted for. An influential paper along these lines is \cite{Witten:1981me}, which also pointed out one of the most serious difficulties one encounters on this path - obtaining chiral fermions. We will further comment on pros and cons of the Kaluza-Klein approach later. Kaluza-Klein scenario is now part of string theory in the sense that the low energy limit of string theory (or hypothetical M-theory) is 10D or 11D supergravity, which then gives gravity plus a variety of matter fields upon compactification to 4D, see e.g. \cite{Duff:1986hr} for a review. 

In the second category one introduces an independent connection field into the Lagrangian of gravity. This can be the affine connection of the Palatini formulation of GR, or the spin connection of Cartan's tetrad formulation, or the self-dual part of the spin connection in Plebanski-related formulations, or the Poincare connection in MacDowell-Mansouri-type formulations. 
In all these formulations (apart possibly from
the MacDowell-Mansouri one), the connection is first introduced as an auxiliary object, to convert the Lagrangian of the theory from second-order in derivatives to first order. In this respect the connection is analogous to an independent momentum variable that can be introduced to convert Lagrangian mechanics to Hamiltonian form. 

Once a connection appears in the Lagrangian, one can also change the viewpoint and think about the connection field, not the metric, as the ``main'' variable. 
This leads to formulations of the third category, in which the metric variable of the first-order formalism is ``integrated out'' to obtain a second-order ``pure connection'' formulation. This only works when there is a non-zero cosmological constant. 

The second and third category offer an alternative route
to unification: instead of enlarging spacetime
one can enlarge the gauge group and the corresponding connection.
This turns out to be much closer in spirit to
unification as normally understood in particle physics.

It is worth mentioning that also this route had been briefly explored by Einstein with his assistant Walther Mayer 
\footnote{better known for the 
Mayer-Vietoris sequence in topology.} 
around 1931 \cite{EM1}, \cite{EM2}.
These papers seem to be very little known,
and we shall review them briefly in section \ref{sec:em}.
A non-abelian generalization of this work, due to Rosen
(another erstwhile collaborator of Einstein) and Tauber appeared in 1984 \cite{rosentauber}.
At about the same time, the same idea was proposed also by 
one of us,
motivated by the analogy to grand unification in particle physics.
In the same year, however, particle physics took a different direction.
This, and the disappearance of Einstein's old school,
is one reason why the bibliography on this scenario 
is so limited. We will give references in the main text. 

Let us also point out that there are many excellent reviews specifically devoted to the Kaluza-Klein scenario, but there is very little discussion in the literature on the possibility of ``enlarging the internal gauge group'', and this is why our review is mostly devoted to the latter scenario.

This paper is organised as follows. We start, in Section \ref{sec:PP-unific}, with a concise description of what unification means in particle physics. We also describe here some less conventional unification scenarios that view the Higgs field as the component of a gauge field in a higher dimensional space. We then review, in Section \ref{sec:formulations}, the known formulations of GR that are relevant for the program of unification. We cover the usual metric formulation (briefly), the tetrad formulation and its versions, the MacDowell-Mansouri formulation and its versions, and BF-type formulations in their non-chiral and chiral forms. Section \ref{sec:hints} describes some hints that support the idea of a classical gravity - YM unification. The various unification scenarios are then treated in Section \ref{sec:unific}. 
Section \ref{sec:disc} contains a critical assessment
of these theories, a discussion of various related topics
and conclusions.

\section{What is unification?}
\label{sec:PP-unific}

In very broad terms unification is a mechanism by which fields of two (or several) different physical theories become components of a single field. There are many examples of this sort in physics.
Here we review the main points and examples in a non-gravitational
context. 

\subsection{Unification in particle physics}
\label{sec:whatis}

In particle physics the notion of unification has a well-defined technical meaning,
that we will use as a benchmark to judge our tentative unified theories including gravity.

Suppose we have two types of interactions $A$ and $B$ described by
two gauge theories with gauge groups $G_A$ and $G_B$.
In order to construct a unified theory of $A$ and $B$
one has to go through the following ``to do'' list.

\begin{enumerate}
\item identify a unifying group $G$ that contains $G_A$ and $G_B$ as commuting subgroups.
\item fit all known particles in irreducible representations of $G$
in such a way that when they are decomposed in
representations of $G_A\times G_B$, all the known particles
are present with the right quantum numbers.
The decomposition will generally 
contain also new particles that are not
in the known low-energy spectrum, for example gauge fields
that are not in the subalgebras of $G_A$ or $G_B$.
\item identify a suitable order parameter. 
Typically, this is a scalar carrying a linear representation of $G$ 
that contains an orbit diffeomorphic to the coset $G/(G_A\times G_B)$.
In general the order parameter need not be a scalar, nor a fundamental field.
\item write a $G$-invariant action for all the fields.
Among other things, the action must be such that:
\begin{enumerate}
\item it should provide a dynamical explanation for the different appearance of the phenomena $A$ and $B$.
This can be obtained by writing a $G$-invariant potential for the
order parameter, whose minima form an orbit with
stabilizer $G_A\times G_B$.
By tuning some of the parameters in the potential (typically a mass),
one can continuously go from an ``unbroken'' or ``unified'' phase,
when the minimum of the potential is in the origin,
to a ``broken'' or ``Higgs'' phase, when the minima of the potential
form an orbit diffeomorphic to the coset $G/(G_A\times G_B)$.
If the order parameter is a composite object, 
a more elaborate dynamical explanation may be possible.
\item all the new particles not contained in the original theories 
must have high mass, so as to be invisible at low energy.
\end{enumerate}
\end{enumerate}

Note that the first three points are of a group-theoretic
or kinematical character, whereas point 4 concerns the dynamics.
One could add to the preceding list some
further requirements, for example that the
theory be renormalizable.
This, however, is not strictly necessary.
We shall return to this point.

In order to have a genuine unified theory the group $G$ should be simple. Then, there is only one gauge coupling and the
difference between the interactions $A$ and $B$ is entirely due to
the non-zero VEV of the order parameter. 
Even though the electroweak (EW) sector of the SM
is not a genuine unified theory, because the EW gauge group $SU(2)\times U(1)$ is not semi-simple, it satisfies all other requirements listed above.
Much work has gone towards a construction of ``Grand Unified Theories''
(GUTs) of the EW and strong interactions \cite{Langacker:1980js}.
In these theories the connection has the block structure
\be
\label{GUT}
\text{GUT GAUGE FIELD} = \left( \begin{array}{cc} \text{EW gauge field} & \text{mixed gauge field} 
\\ 
\text{mixed gauge field} & \text{strong gauge field} \end{array} \right)\ .
\ee
where the mixed gauge field components (usually called ``leptoquarks'') must have a very high mass.
There are many possible variants, both with and without supersymmetry.
The lack of progress on this front is due mainly to the
failure to detect the decay of the proton, which is predicted by
these models.
Many models have been ruled out but others remain viable
\cite{Bertolini:2009qj,Altarelli:2013aqa}.

In the preceding ``to do'' list we have described the case
when the order parameter carries a linear representation.
This is the case, e.g. for the EW theory, where the order parameter is the Higgs field, which carries the fundamental spinor representation of the weak $SU(2)$. 
One can also go through the same steps by assuming that the
order parameter carries a nonlinear realization.
In this case it would be typically a scalar
with values in the coset $G/(G_A\times G_B)$
(a nonlinear sigma model).
Thus, we would have only a subset of the fields that are present
in the linear models, namely those variables that can be
called ``Goldstone bosons''.
Insofar as the purpose of the above construction is to
give a mass to the components of the gauge fields that are
not in the subalgebras of $G_A$ or $G_B$,
this is perfectly sufficient:
one can choose a ``unitary'' gauge where the Goldstone boson
fields are constant, and in this gauge the kinetic term
of the scalars becomes a mass term for some components
of the gauge field \cite{appel,longhitano,herrero}.
However, there are several reasons why this is not completely satisfactory.
First, the nonlinear sigma model is not renormalizable.
Second, and even more important,
the scattering amplitude of the longitudinal bosons
would violate the unitarity bounds near the scale of unification.
This is avoided in the presence of the additional singlet 
``Higgs'' scalar
degrees of freedom that form a linear representation.
Finally, the non-linearly-realized theory describes only the
``broken'' phase.
For all these reasons, at least in the particle physics context, the non-linearly realized theory
should only be viewed as an effective field theory
valid at energies below the scale of unification
(the coupling of the nonlinear sigma model is roughly
comparable to the inverse of the unification scale). 


\subsection{Gauge-Higgs unification}

Apart from the standard particle physics unification scheme reviewed above, there is a set of scenarios where further unification is achieved. 
There are several different realisations, but they all share the same common idea --- it is assumed that the space(time) has extra dimensions and the Higgs fields are just the extra-dimensional components of the gauge field. In other words, a pure gauge theory in higher number of dimensions gives rise to a gauge theory plus scalar (Higgs) fields upon dimensional reduction to lower dimensions:
\be
\label{gaugehiggs}
\text{GAUGE FIELD} = \left( \begin{array}{c} \text{Higgs}\\ 
\text{gauge field} \end{array} \right)\ .
\ee
This is in the spirit of KK dimensional reduction, the only difference being that the theory that is reduced is Yang-Mills theory in higher dimensions, not gravity. 

At the mathematical level, there are some famous realisations of this mechanism. $N=4$ supersymmetric Yang-Mills theory in four dimensions, with its six (Lie algebra valued) scalars arises as the dimensional reduction of the ten-dimensional Super-Yang-Mills theory \cite{Brink:1976bc}. Hitchin integrable system in two dimensions with its complex-valued Higgs field arises as the dimensional reduction of the four-dimensional self-dual Yang-Mills theory \cite{Hitchin:1986vp}.

From a more physical perspective, the interpretation of the Higgs field as the extra-dimensional component of the gauge field has been developed by Hosotani, see e.g. \cite{Hosotani:2017krs} and references therein. The non-commutative geometry model of Connes interprets the bosonic sector of EW theory as the Yang-Mills theory on a non-commutative space \cite{Connes:1990qp}. The Higgs field gets the interpretation of the component of the gauge field along the non-commutative direction. A similar idea is used in the approach pioneered by Neeman, where the Higgs field appears as a component of a superconnection, see \cite{Neeman:2005dbs} for a review, and \cite{Aydemir:2014ama} for a recent concrete scenario along these lines. 

To summarise, apart from standard particle physics models that require dedicated Higgs fields to break symmetry, there are also scenarios that give Kaluza-Klein-type interpretation to the Higgs fields, as extra-dimensional components of a gauge field. There is a host of models along these lines, and the difficulty is always in writing a realistic model. 

\goodbreak

\section{Formulations of General Relativity}
\label{sec:formulations}

As we have described in the Introduction, in very general terms the idea of unifying gravity with other forces is to take a particular formulation of General Relativity and ``enlarge'' the fields appearing in the corresponding Lagrangian. The unification procedure will thus depend on the formulation of GR that one takes as the starting point. Because of this, we will start by reviewing the available formulations.

\subsection{The metric-related formulations}

\subsubsection{Metric formulation}

This is standard, so we will be very brief. The only field appearing in the action is (at this stage symmetric) metric. The Einstein-Hilbert action in $D$ spacetime dimensions is
\be
\label{action-EH}
S_{\rm EH}[g] = \frac{1}{16\pi G} \int d^Dx \sqrt{-g} \left( R- (D-2)\Lambda \right),
\ee
where $G$ is the Newton's constant. In $D$ dimensions it has the mass dimension $[1/G_N]=D-2$, while $[\Lambda]=2$.
The cosmological constant is normalised so that,
in any dimension, the Einstein equation in the absence of matter
reads $R_{\mu\nu}=\Lambda g_{\mu\nu}$. The sign in front of the action
is signature- and convention-dependent, see below for ours. 

\subsubsection{First-order Palatini formulation}
\label{sub-sec:palatini}

In the first-order formulation one introduces an independent connection field into the game, to convert the Lagrangian into first order in derivatives form. The Lagrangian is
\be\label{action-palatini}
S_{\rm Palatini}[g,\Gamma] = \frac{1}{16\pi G}\int d^Dx \sqrt{-g}  \left(  g^{\mu\nu} R_{\mu\nu}(\Gamma) - (D-2)\Lambda\right).
\ee
Here $\Gamma_\mu{}^\rho{}_\nu$ is the affine connection, i.e. a connection on the tangent bundle to a manifold, with the covariant derivative being $\nabla_\mu v^\nu = \partial_\mu v_\nu + \Gamma_\mu{}^\rho{}_\nu v^\rho$. Our convention for the Riemann curvature is $- 2\nabla_{[\mu}\nabla_{\nu]} v^\rho =  R^\rho{}_{\sigma\mu\nu} v^\sigma$ so that
\be
\label{riemanncurv}
R^\sigma{}_{\rho\mu\nu} = 
\partial_\nu \Gamma_\mu{}^\sigma{}_{\rho} 
- \partial_\mu \Gamma_\nu{}^\sigma{}_{\rho} 
+  \Gamma_\nu{}^\sigma{}_\alpha \Gamma_\mu{}^\alpha{}_{\rho}
-  \Gamma_\mu{}^\sigma{}_{\alpha} \Gamma_\nu{}^\alpha{}_{\rho}\ .
\ee
One forms the Ricci tensor present in (\ref{action-palatini}) out of the Riemann curvature $R_{\mu\nu}(\Gamma):= R^\sigma{}_{\mu\sigma\nu}$. In Palatini formalism the affine connection is assumed to be torsion-free, i.e. to satisy the symmetry
\footnote{Actually one could also assume that the connection
is metric-compatible and derive the absence of torsion
as an equation of motion. See section \ref{sub-sec:gln}  below.}
\be
\label{symconn}
\Gamma_\nu{}^\mu{}_{\rho}=\Gamma_\rho{}^\mu{}_{\nu}\ . 
\ee

The Ricci curvature $R_{\mu\nu}(\Gamma)$ is not automatically symmetric, but the symmetric part is selected in (\ref{action-palatini}) when $R_{\mu\nu}$ gets contracted with the symmetric metric.  

Variation of (\ref{action-palatini}) with respect to the affine connection gives an equation that implies that $\nabla_\rho g^{\mu\nu}=0$, i.e. that the connection is metric-compatible. The solution to this equation is the usual expression for $\Gamma$ in terms of the derivatives of the metric. Substituting this solution into the action one gets back the second-order Einstein-Hilbert action (\ref{action-EH}). 

We also note that in the case $\Lambda=0$, if one views $\sqrt{-g} g^{\mu\nu}$ as the basic variable of the theory, the action (\ref{action-palatini}) is cubic in the fields. This has been emphasised by Deser \cite{Deser:1969wk}, who used this cubic formulation to reconstruct GR from the linear Fierz-Pauli theory and hence prove its uniqueness. 

\subsubsection{Eddington-Schr\"odinger formulation}

Instead of ``integrating out'' from (\ref{action-palatini}) the affine connection to get back (\ref{action-EH}) one can integrate out the metric field. Indeed, varying the Palatini action with respect to the metric one gets an equation that is trivially solved 
\be\label{metric-ES}
g_{\mu\nu} = \frac{1}{\Lambda} R_{(\mu\nu)}(\Gamma).
\ee
This can then be substituted into the action to get a second-order pure affine formulation
\be\label{action-ES}
S_{\rm ES}[\Gamma] = \frac{1}{8\pi G \Lambda^{(D-2)/2}} \int d^Dx \sqrt{ - {\rm det}(R_{(\mu\nu)}(\Gamma))}.
\ee
The field equation that results by varying this action with respect to the connection implies that the metric defined in (\ref{metric-ES}) is compatible with the connection. The definition of the metric (\ref{metric-ES}) then becomes the Einstein equation. We note that this purely affine formulation is only available with a non-zero cosmological constant. Note also that the coefficient in front of the Eddington-Schr\"odinger action is always dimensionless. In four dimensions we have $(G\Lambda)^{-1} \sim 10^{120}$, a very large number.

While the action (\ref{action-ES}) appears to be a natural construct, the pure affine formalism brings with it arbitrariness that is not present in the metric formalism. This has been emphasised in particular by Pauli, see \cite{Goenner:2014mka}, Section 8.2. Thus, the tensor $R_{\mu\nu}$ is not automatically symmetric even for a symmetric affine connection. It can be split into its symmetric and anti-symmetric parts, and these can be separately used in constructing the Lagrangian. The elementary building blocks are then 
\be
L_0= \sqrt{ - {\rm det}(R_{(\mu\nu)}(\Gamma))}, \quad 
L_1 = \sqrt{ \tilde{\epsilon}^{\mu\nu\rho\sigma} \tilde{\epsilon}^{\alpha\beta\gamma\delta} R_{(\mu\alpha)} R_{(\nu\beta)} R_{[\rho\gamma]} R_{[\sigma\delta]}}, \quad
L_2 = \tilde{\epsilon}^{\mu\nu\rho\sigma} R_{[\mu\nu]} R_{[\rho\sigma]},
\ee
where $\tilde{\epsilon}^{\mu\nu\rho\sigma}$ is the densitized anti-symmetric tensor that exists without any background structure on the manifold. The above blocks are all densities of weight one, and can be integrated over the manifold. However, one can also consider their ratios. The most general Lagrangian is then 
\be
L = L_0\, f\left( \frac{L_1}{L_0}, \frac{L_2}{L_0}\right)
\ee
for an arbitrary function $f$ of 2 variables. The case $f=1$ gives GR, but other choices are possible. A general theory from this class has been studied in \cite{Hejna:2006hd}, where it was shown that it is equivalent to a non-linear Einstein-Proca system. This ambiguity in writing down the most general Lagrangian is a drawback of all ``pure connection'' formulations, see below.

Another drawback of the pure affine formulation is the very large number of field components one has to deal with. 
Indeed, in four dimensions we have $4\times 10=40$ components in $\Gamma_\mu{}^\rho{}_{\nu}$ as compared to only $10$ components in $g_{\mu\nu}$. This makes the pure affine formalism not too useful in practice. 

\subsection{Tetrad and related formulations}

\subsubsection{Frame fields and their geometrical interpretation}

The tetrad (a.k.a. vierbein, or co-frame field
\footnote{The terms ``tetrad'' and ``'vierbein'' 
both have the drawback that they
refer explicitly to four dimensions.
In three dimensions the same fields are usually referred to
as triads or dreibeins; in higher dimensions the term ``vielbein''
is used. We will ignore this and use the same term in all dimensions.}
) 
is a collection of $D$ linearly independent one-forms $\theta^I{}_\mu$ such that
\be
\label{tet}
g_{\mu\nu}=\theta^I{}_\mu\theta^J{}_\nu \eta_{IJ}\ .
\ee 
An equivalent set of variables is given by the frame field
(sometimes also called the inverse tetrad or inverse vierbein)
$\theta_I{}^\mu$, which is a collection of $D$ linearly independent 
vectors.
They are related by
\be
\theta^I{}_\nu \theta_I{}^\mu=\delta^\mu_\nu\ ;\qquad
\theta^I{}_\mu \theta_J{}^\mu=\delta^I_J\ .
\ee
The geometrical interpretation of $\theta_I{}^\mu$
is as fields of orthonormal frames in the tangent bundle of spacetime, and of $\theta_\mu^I$ as orthonormal co-frames in the cotangent bundle.
Then, every tensor can be decomposed in such frames, for example
the orthonormal components of a tensor $t$ are related to the
components in a coordinate basis by
$t^{IJ}_{KL}=\theta^I{}_\mu\theta^J{}_\nu
\theta_K{}^\alpha\theta_L{}^\beta t^{\mu\nu}_{\alpha\beta}\ .$
Also the connection can be written in the orthonormal basis.
The orthonormal components of the connection,
denoted $\omega_\rho{}^I{}_J$,
are related to the Christoffel symbols,
$\Gamma_\rho{}^\mu{}_{\nu}$ 
({\it i.e.} the coordinate components of the Levi-Civita connection) by
\footnote{It would be more natural to denote these
by $\Gamma_\rho{}^I{}_J$, because they are the components
of the same connection in a different basis.
We will stick to the traditional notation.}
\be
\label{spinconn}
\partial_\rho{}\theta^I{}_\mu
+\omega_\rho{}^I{}_J\theta^J{}_\mu
-\theta^I{}_\sigma\Gamma_\rho{}^\sigma{}_{\mu}
=0\ .
\ee
This equation is usually interpreted as saying that
``the total covariant derivative of the tetrad vanishes''
and written in the form $\nabla_\rho \theta^I{}_\mu=0$. 
When $\Gamma_\rho{}^\nu{}_{\mu}$ is compatible with the spacetime metric, in the sense that $\nabla_\rho g_{\mu\nu}=0$, the connection $\omega_\rho{}^I{}_J$ is compatible with the internal metric $\eta^{IJ}$. Indeed, we have $\eta^{IJ}=\theta^I{}_\mu \theta^J{}_\nu g^{\mu\nu}$ and so $\nabla_\rho \eta^{IJ}=0$ because both the covariant derivative of the tetrad is zero (\ref{spinconn}), and the covariant derivative of the spacetime metric is zero. 
The statement that the connection
is compatible with the metric $\eta_{IJ}$ is the statement 
that it is a Lorentz connection. This connection is also referred to as the spin connection.
\footnote{The Spin group is the double cover of the Lorentz group. 
These groups have the same Lie algebra and therefore a connection for one is also a connection for the other.}
Note that, defining $\omega_{\rho IJ}=\eta_{IK}\omega_\rho{}^K{}_J$, the metricity of $\omega$ is just the condition of anti-symmetry 
\be
\label{antisymmconn}
\omega_{\mu IJ}=-\omega_{\mu JI}\ .
\ee

It is clear that for a given metric, 
the orthonormal frame is not unique 
-- the Lorentz rotated co-frame
\be
\theta^{\prime I}{}_\mu=\Lambda^{-1 I}{}_J\theta^J{}_\mu
\ee 
gives the same metric. 
This brings the group of local Lorentz rotations into play. This is an important point for later, because the unification procedure that we will consider below will consist in enlarging this group. 
Under local Lorentz transformations the spin connection transforms as
\be
\omega'_\mu=\Lambda^{-1}\omega_\mu\Lambda
+\Lambda^{-1}\partial_\mu\Lambda
\ee
where we treated $\omega_\mu$ as a matrix, suppressing the
Lorentz indices.
In the same notation, the curvature of the spin connection is
\be
R_{\mu\nu}=-\partial_\mu\omega_\nu+\partial_\nu\omega_\mu
-[\omega_\mu,\omega_\nu]
\ee
and is related to the Riemann tensor by 
$R_{\mu\nu IJ}=\theta_I{}^\rho\theta_J{}^\sigma R_{\mu\nu\rho\sigma}$.
(The choice of sign is dictated by consistency
with equation (\ref{riemanncurv}).)

Given any action $S(g)$ for gravity in metric formulation,
one obtains an action in the tetrad formulation
by setting $S'(\theta)=S(g(\theta))$,
where $g(\theta)$ is given by (\ref{tet}).
For example starting from the Hilbert action (\ref{action-EH}):
\be
\label{tetac}
S'(\theta)=\frac{1}{16\pi G}\int d^Dx|\det\theta|\,
\theta_I{}^\mu\theta_J{}^\nu
R_{\mu\nu}{}^{IJ}\ .
\ee

The tetrads are necessary to couple gravity to spinor fields, 
because spinors are representations of the Spin group.
One cannot write Dirac Lagrangian for the fermions in terms of the metric alone.

The interpretation of the frame fields given above is the most straightforward one but it has the drawback that a smooth assignment of frame fields is in general only possible locally. 
It is somewhat unusual that the dynamical variable should not be
a globally defined geometrical object.

There is an alternative interpretation that does not have this drawback.
One can think of a vector bundle $E$ with fibers $\R^D$ 
and a fiber metric of desired signature, that is globally isomorphic
to the tangent bundle.
Let $\theta$ be an isomorphism of $TM$ to $E$.
We choose (locally) orthonormal frames $\{e_I\}$ in $E$ and stick to
coordinate frames $\{\partial_\mu\}$ in $TM$.
Then we can view $\theta^I{}_\mu$ as the local matrix representation 
of the isomorphism, relative to these bases.
In this interpretation, the co-frame field $\theta^I{}_\mu$
is also called the ``soldering form''.
Equation (\ref{tet}) says that the metric $g$ 
on spacetime is the pullback by $\theta$ of the fiber metric in $E$
and likewise (\ref{spinconn}) expresses the connection in $TM$
as the pullback of the connection in $E$.
In this interpretation, $E$ is a priori unrelated to spacetime
and therefore its fibers can be thought of as ``internal'' spaces.

Throughout the rest of the paper we shall implicitly adopt
this second interpretation and refer to latin indices $I,J,\ldots$ as ``internal indices''.

\subsubsection{Einstein-Cartan formulation}
\label{sub-sec:EC}

The Einstein-Cartan formulation is first-order in derivatives, so that apart from the tetrad there is also an independent Lorentz connection 
$\omega_{\mu IJ}$. On-shell it becomes related to the connection in $TM$ by (\ref{spinconn}). Importantly, the Einstein-Cartan action is written in terms of differential forms and their wedge products, which makes it, unlike its Einstein-Hilbert counterpart, polynomial. We will only need the 4D version. The action reads
\be\label{action-EC}
S_{\rm EC}[\theta,\omega] = \frac{1}{32\pi G} \int \epsilon_{IJKL} \, \theta^I\wedge \theta^J \wedge \left( F^{KL}(\omega) - \frac{\Lambda}{6} \theta^K\wedge \theta^L\right).
\ee
Here $F^{IJ}(\omega) = d \omega^{IJ} + \omega^{IK}\wedge \omega_K{}^J$ is the curvature of the Lorentz connection.

When one varies (\ref{action-EC}) with respect to the connection, one obtains an equation that implies $\nabla \theta^I \equiv d\theta^I + \omega^I{}_J \wedge \theta^J=0$, i.e. the zero torsion condition. This is an algebraic equation for $\omega^{IJ}$, and can be solved uniquely in terms of the derivatives of $\theta^I$. Substituting this solution into the action (\ref{action-EC}) brings us back to the
action (\ref{tetac}) discussed in the preceding subsection. 

We also note that the tetrad $\theta^I$ and spin connection $\omega^{IJ}$ are differential forms. Given a metric, this gives a very efficient way of computing the Riemann curvature. This is in particular due to the fact that in 4D there are only $4\times 6=24$ components of $\omega^{IJ}$ to solve for, while the affine connection $\Gamma_\mu{}^\rho{}_{\nu}$ has $4\times 10$ components. Once the spin connection is known the Levi-Civita connection is recovered via (\ref{spinconn}).

We note that the Einstein-Cartan action (\ref{action-EC}) is polynomial in the fields it contains, and contains just up to quartic terms. This is true even for $\Lambda\not=0$, in contrast to the case of the Palatini action (\ref{action-palatini}) which is only polynomial (with the choice of the inverse densitiesed metric as the main variable) for $\Lambda=0$. This, as well as the necessity of tetrads when spinors are present, are the two reasons why the tetrad formulation can be considered superior to the formulation in terms of the metric. 

One drawback of the Einstein-Cartan formulation as compared to the metric one is more complicated character of its Hamiltonian formulation obtained via the $3+1$ split. It is known that in this case there are second class constraints, see e.g. \cite{Holst:1995pc} for the Hamiltonian analysis. This should be contrasted with the ADM formalism \cite{Arnowitt:1960es} where no second class constraints appear. The appearance of second class constraints in the Einstein-Cartan formalism is not surprising because 24 ``momentum'' variables have been introduced in addition to 16 ``configuration'' variables. The extra variables are then eliminated by second class constraints. A formalism that shares all the good features of Einstein-Cartan but does not suffer from the problem of second class constraints is the chiral first-order formalism to be reviewed below. 

\subsubsection{Pure Lorentz connection formulation}

Given that it is possible to ``integrate out'' the metric variable from Palatini Lagrangian (\ref{action-palatini}) to obtain the pure affine formulation (\ref{action-ES}), one can ask whether a similar trick is possible with the Einstein-Cartan formulation. The field equations one gets for the tetrad are algebraic in any dimension, so this is always possible in principle. In 3D it is possible to obtain a closed form expression for the corresponding pure connection Lagrangian, see \cite{Peldan:1991mh} and also \cite{Herfray:2016std} for the description of this functional. In 4D the equation one needs to solve is
\be\label{tetrad-ee}
\epsilon_{IJKL} \theta^J \wedge F^{KL} = \frac{\Lambda}{3} \epsilon_{IJKL} \theta^J \wedge \theta^K\wedge \theta^L.
\ee
At present it is not known how to solve this equation for $\theta^I$ in a closed form. However, a perturbative solution (around constant curvature background) is possible, see \cite{Zinoviev:2005qp,Basile:2015jjd}.

We now describe this solution. The constant curvature background corresponds to 
\be\label{pure-spin-backgr}
F^{IJ}= \frac{\Lambda}{3} \theta^I\wedge \theta^J.
\ee
Denoting by $\theta^I, \omega^{IJ}$ the background and by $e^I, a^{IJ}$ the perturbations we have the following linearisation of (\ref{tetrad-ee})
\be
\epsilon_{IJKL} \theta^J \wedge \nabla a^{KL} = \frac{2\Lambda}{3} \epsilon_{IJKL} e^J \wedge \theta^K\wedge \theta^L,
\ee
whose solution is
\be\label{e-pure-spin-conn}
e^I= \frac{3}{2\Lambda} \hat{R}^I_J \theta^J, \qquad \hat{R}^I_J := R^I_J - \frac{1}{6} \delta^I_J R,
\ee
where we introduced the linearised curvature $R^{IJ}_{KL}:=2 \nabla_{[\mu} a^{IJ}_{\nu]} \theta^\mu_K \theta^\nu_L $ and  $R^{I}_J=  R^{IK}_{JK}, R= R^I_I$. Note that the linearised ``Ricci'' tensor $R^I_J$ does not need to be symmetric. 

The linearisation of the action (\ref{action-EC}), evaluated on the solution (\ref{e-pure-spin-conn}) gives, compare \cite{Zinoviev:2005qp}
\be\label{pure-spin-action-1}
S^{(2)}[a] = \frac{3}{32\pi G\Lambda} \int \theta (\delta^I_K \delta^J_M - \delta^I_M \delta^J_K) \hat{R}^K_I \hat{R}^M_J + \frac{\Lambda}{3} \epsilon_{IJKL} \theta^I \wedge \theta^J \wedge a^K{}_M \wedge a^{ML},
\ee
where $\theta:=(1/24)\epsilon_{IJKL} \theta^I\theta^J\theta^K\theta^L$ is the volume form for $\theta^I$. The last term here can be rewritten in a convenient form. Thus, one uses the background condition (\ref{pure-spin-backgr}) to replace the wedge product of two $\theta$'s with the curvature. The term $\epsilon_{IJKL} F^{IJ} a^K{}_M a^{ML}$ is then rewritten by replacing $a^{ML} = (1/4) \epsilon^{MLPQ} \epsilon_{PQRS} a^{RS}$, and decomposing the product of two of the $\epsilon$'s. We get
\be
\epsilon_{IJKL} F^{IJ} a^K{}_M a^{ML} = F^{IM} a_M{}^J \epsilon_{IJKL} a^{KL} = (1/2) (\nabla\nabla) a^{IJ} \epsilon_{IJKL} a^{KL}.
\ee
Integrating by parts we can then replace the last term in (\ref{pure-spin-action-1}) with $-\epsilon_{IJKL} \nabla a^{IJ} \nabla a^{KL}= - (\theta/4) \epsilon_{IJKL} \epsilon^{MNPQ} R^{IJ}_{MN} R^{KL}_{PQ}$. In these manipulations the wedge product is implied everywhere. Thus, the last term in (\ref{pure-spin-action-1}) can also be rewritten in the form curvature squared. The final result for the linearised action can be written very compactly as \cite{Basile:2015jjd}
\be\label{pure-spin-action-2}
S^{(2)}[a] = - \frac{3}{64\pi G\Lambda} \int \theta \, C_{IJ}^{KL}[a] C_{KL}^{IJ}[a],
\ee
where the Weyl-like tensor is defined as
\be
C^{IJ}_{KL}[a] := R^{IJ}_{KL} - (\delta^I_{[K} R^J_{L]} - \delta^J_{[K} R^I_{L]}) + \frac{R}{3} \delta^I_{[K} \delta^J_{L]}.
\ee
Note that in Euclidean signature the action (\ref{pure-spin-action-2}) has a definite sign. This is similar to Eddington-Schr\"odinger action (\ref{action-ES}), but in contrast to the metric formulation (\ref{action-EH}). The above manipulations can be simplified by starting with the MacDowell-Mansouri action instead, as in \cite{Basile:2015jjd}. In that case there is no need for integration by parts manipulations, and the linearised action (\ref{pure-spin-action-2}) results immediately. We will review this below. 

\subsubsection{GL(D) formalism}
\label{sub-sec:gln}

In section \ref{sub-sec:palatini} we worked with fixed coordinate frames, used the metric and the affine connections as the basic fields and imposed absence of torsion on the dynamical connection, which translates into the purely algebraic symmetry condition (\ref{symconn}).
Dynamics then enforced metric-compatibility.
On the other hand in section \ref{sub-sec:EC} we worked with dynamical orthonormal frames, so that the components of the metric were fixed, and imposed metricity on the connection, which translates into the purely algebraic antisymmetry condition (\ref{antisymmconn}).
Dynamics then enforced the absence of torsion.

One may wonder whether one could have exchanged the roles of
torsion and non-metricity.
The two formulations only differ in the choice of frames,
and physics cannot depend on such a choice, so one would expect
the answer to be positive.
This is indeed the case, and to make it manifest one can use a more general formulation, where the frames are not restricted to be either natural or orthonormal, and the connection is not constrained a priori to satisfy any condition.
Then, equation (\ref{tet}) is generalized to
\be
\label{gl4met}
g_{\mu\nu}=\theta^I{}_\mu\theta^J{}_\nu \gamma_{IJ}\ .
\ee 
where $\gamma_{IJ}$, a set of scalar fields, are the components
of the metric in the vectorbundle $E$ and the connection in $TM$
is
\be
\label{gl4conn}
\Gamma_\rho{}^\mu{}_{\nu}=
\theta_I{}^\mu\omega_\rho{}^I{}_J\theta^J{}_\nu
+\theta_I{}^\mu\partial_\rho\theta^I{}_\nu
\ee
where $\omega_\mu$ now does not have any symmetry property.
In a general gauge, torsion and non-metricity both
involve derivatives:
\bea
\label{torsion}
\Theta_\mu{}^I{}_\nu&=&
\partial_\mu\theta^I{}_\nu
-\partial_\nu\theta^I{}_\mu
+\omega_\mu{}^I{}_J\theta^J{}_\nu
-\omega_\nu{}^I{}_J\theta^J{}_\mu\ ,
\\
\label{nonmet}
\Delta_{\mu IJ}&=&-\partial_\mu\gamma_{IJ}
+\omega_\mu{}^K{}_I\gamma_{KJ}
+\omega_\mu{}^K{}_J\gamma_{IK}\ .
\eea
In such a formulation one is free to perform local linear transformations on the indices $I$, $J$, so the local Lorentz-invariance of the tetrad formulation is extended to local $GL(D)$-invariance. 
This invariance can be gauge-fixed by either fixing the soldering form $\theta^I_\mu=\delta^I_\mu$, which brings us back to the standard formulation in natural frames, or the fiber metric $\gamma_{IJ}=\eta_{IJ}$, which leads to the vierbein formulation.
Note again that torsion is purely algebraic in the first gauge and
non-metricity is algebraic in the second one.

One can write an action 
\be\label{action-gln}
S(\theta,\gamma,\omega)=S_{\rm Palatini}(g(\theta,\gamma),\Gamma(\theta,\omega)),
\ee
where the metric $g$ and connection $\Gamma$ are given by equations (\ref{gl4met}),(\ref{gl4conn}).
It turns out that in this generalized context the variation with respect to $\omega$ does not fix the connection uniquely. This is due to the projective invariance of the action, namely invariance under the transformations
$\delta\omega_\mu{}^I{}_J=\delta^I_J v_\mu$.
One can get around this by demanding either metricity or torsionlessness, and then the other follows from the equations of motion. 
However, the condition to be imposed is now a differential,
not a purely algebraic one as in the Palatini or Einstein-Cartan
formulations.
Alternatively, we can further modify the action by adding a generic term involving the squares of $\Theta$ and $\Delta$, which can be seen
as the natural kinetic terms of the fields $\theta$ and $\gamma$.
One can show that generically (i.e. for almost all choices of coefficients of such terms) the field equations imply that the connection is metric and torsion-free, and on shell the theory is equivalent to GR. 

Note that the action (\ref{action-gln}) is no longer first order in derivatives, as the affine connection $\Gamma_\rho{}^\mu{}_{\nu}$ now contains a derivative of the frame. 
One of the reasons for introducing an independent connection
was the desire to have an action that is first order in derivatives.
From this point of view, the action (\ref{action-gln}),
possibly supplemented by terms quadratic in torsion and non-metricity,
could be seen as a step back. 
A related point is that there appears to be no way to write an action realising these ideas in terms of differential forms, which can also be 
seen as a drawback of this formalism. 

Once the kinetic terms for the frame and metric are introduced as suggested above, one can see that there is a kind of Higgs mechanism in action, giving mass to the connection, or more precisely to the difference of the dynamical connection from the Levi-Civita connection. This  effectively removes the connection from the low energy spectrum, independently of the details of the gravitational dynamics, and further strengthens the interpretation of GR as a low-energy effective field theory.

This formalism has various applications \cite{tmg,anomaly,siegel}.
It is necessary for a proper understanding of the transformation of spinors
under diffeomorphisms \cite{dabrowski}.
The $GL(D)$-invariant connection can be coupled to spinors 
by first extracting from it a Lorentz connection. 
This is possible and unambiguous in presence of $\theta$ and $\gamma$. However, the interpretation of the remaining, non-metric degrees of freedom is not very clear and therefore this formulation is not very natural for what we are going to discuss in the following. Also it seems that fermions, and in particular SM fermions to be reviewed below, suggest that the gauge group of the theory should be an orthogonal group, not general linear group. So, this type of generalisation does not appear to go in the right direction. We refer to \cite{perbook1,Percacci:2009ij} for more details on this formalism. 

\subsection{MacDowell-Mansouri formulation}

The idea of this formulation \cite{MacDowell:1977jt} is to combine the spin connection $\omega^{IJ}$ of the Einstein-Cartan formalism together with the tetrad $\theta^I$ into a connection for the gauge group
$SO(1,4)$ or $SO(2,3)$, depending on the sign of the cosmological constant. The Lie algebra of these groups splits as the sum of the Lorentz subalgebra plus an additional 4-dimensional part. The frame receives the interpretation of the component of the connection in this 4-dimensional part. A similar idea can be put to use in 3D gravity, where it leads to its Chern-Simons formulation \cite{Witten:1988hc} and, when the cosmological constant is zero, in Poincar\'e gauge theories of gravity, see e.g. \cite{Hehl:2012pi}. 

There are two versions of this formulation. In the original
formulation of MacDowell and Mansouri \cite{MacDowell:1977jt}, the basic field is an $SO(1,4)$ or $SO(2,3)$ connection, but the Lagrangian is only invariant under the 4-dimensional Lorentz group.\footnote{Supergravity can also be described along the same lines, by replacing the gauge group that gives pure gravity with a supergroup, see \cite{MacDowell:1977jt}.}  Invariance under $SO(1,4)$ or $SO(2,3)$ is explicitly broken.
In the second version \cite{Stelle:1979aj} the symmetry breaking
from $SO(1,4)$ or $SO(2,3)$ to $SO(1,3)$ is dynamical, due to an auxiliary vector field, often referred to as the compensator in the literature. 

\subsubsection{MacDowell-Mansouri version}

The curvature of an $SO(1,4)$ or $SO(2,3)$ connection has two parts. First, there is the part valued in the Lie algebra of the Lorentz group $SO(1,3)$. It is given by
\be\label{curv-MM}
{\cal F}^{IJ} = F^{IJ}(\omega) -\frac{\Lambda}{3} \theta^I\wedge \theta^J.
\ee
Second, there is the remaining part, which is just a multiple of the torsion tensor $\nabla \theta^I$. The 4-dimensional MacDowell-Mansouri action is
\be\label{action-MM}
S_{\rm MM}[\theta,w]= - \frac{3}{64\pi G\Lambda} \int \epsilon_{IJKL} {\cal F}^{IJ} \wedge {\cal F}^{KL}.
\ee
Using (\ref{curv-MM}) we get the Einstein-Cartan action (\ref{action-EC}) plus a topological term. 

The action (\ref{action-MM}) thus differs from (\ref{action-EC}) by a total derivative term, and leads to the same field equations. However, it has many advantages over the Einstein-Cartan action. First, its value on maximally symmetric backgrounds ${\cal F}^{IJ}=0$ is zero. Second, in relation to the problem of evaluating the gravitational action on e.g. asymptotically Anti-de Sitter spaces, the usual Einstein-Hilbert or Einstein-Cartan actions diverge on such backgrounds and require renormalisation. This is usually done by adding to the action appropriate boundary terms that also diverge as one approaches the AdS boundary. The difference between the divergent bulk and boundary actions is then the renormalised action, see e.g. \cite{deHaro:2000vlm}. The action (\ref{action-MM}) vanishes on exact AdS and is finite on asymptotically AdS solutions. Moreover, the difference between the Einstein-Cartan and MacDowell-Mansouri actions is a total derivative, or equivalently a boundary term. Thus, the boundary terms needed for the renormalisation on asymptotically AdS backgrounds are automatically included in (\ref{action-MM}).

Another advantage of (\ref{action-MM}) over (\ref{action-EC}) is that it is very easy to linearise this action on maximally symmetric backgrounds. Indeed, we have
\be\label{action-MM-lin}
S^{(2)}_{\rm MM}[e,a] = - \frac{3}{64\pi G\Lambda} \int \epsilon_{IJKL} \left(\nabla a^{IJ} - \frac{2\Lambda}{3} \theta^I \wedge e^J\right) \wedge \left(\nabla a^{KL} - \frac{2\Lambda}{3} \theta^K \wedge e^L\right),
\ee
where, as in the previous subsection, $e^I, a^{IJ}$ are the perturbations of the tetrad and the spin connection respectively. Substituting here the solution (\ref{e-pure-spin-conn}) gives the pure connection linearised action (\ref{pure-spin-action-2}) with very little work. Indeed, the combination that appears in (\ref{action-MM-lin}) evaluates to
\be
\nabla a^{IJ} - \frac{2\Lambda}{3} \theta^{[I} \wedge e^{J]} = \frac{1}{2} \left( R^{IJ}_{KL} - 2 \delta^{[I}_M \hat{R}^{J]}_N\right) \theta^M\wedge \theta^N = \frac{1}{2} C^{IJ}_{MN}[a] \theta^M\wedge \theta^N,
\ee
and the result (\ref{pure-spin-action-2}) follows immediately. 

In the Mc-Dowell-Mansouri formulation the fields of the first-order formulation (\ref{action-EC}) have been unified into a single connection field, but now the Lagrangian (\ref{action-MM}) is no longer manifestly of first-order. Schematically, it is of the type $F^2$.  However, the two-derivative term in (\ref{action-MM}) is, modulo total derivative terms, a term with no derivatives. This is why (\ref{action-MM}) is equivalent to the first-order Einstein-Cartan Lagrangian. 

A final remark is that it is possible to put (\ref{action-MM}) into a manifestly first order form by ``integrating in'' a 2-form field, as in BF-type formulations that we consider below. This manifestly first order form of MacDowell-Mansouri theory has been studied in \cite{Freidel:2005ak}.

\subsubsection{Stelle-West version}

The action (\ref{action-MM}) can be rewritten in manifestly $SO(1,4)$ or $SO(2,3)$ invariant form by introducing an extra field. Let us denote the 5-dimensional indices by lower case latin letters, so that $SO(1,4)$ or $SO(2,3)$ Lie algebra valued objects are of the form $v^{ab}=v^{[ab]}$. Let us introduce a new field $v^a$. This field is required to have unit norm $|v|^2=\pm 1$, depending on the sign of the cosmological constant. Let us consider the following action
\be\label{action-SW}
S[A,v]=- \frac{3}{64\pi G\Lambda} \int \epsilon_{abcde} {\cal F}^{ab}(A) {\cal F}^{cd}(A) v^e.
\ee
Here $A^{ab}$ is a $SO(1,4)$ or $SO(2,3)$ connection, and ${\cal F}^{ab}(A)$ is its curvature. The action is manifestly invariant under the large group. Choosing $v^a$ to point in a particular direction breaks the symmetry down to the Lorentz group, and reproduces (\ref{action-MM}). The unit norm constraint can be explicitly added to the action with a Lagrange multiplier, see below. 

To couple gravity in this form to matter one just has to note that the frame is readily recovered as the covariant derivative $\nabla v^a$ (with respect to the connection $A^{ab}$) of the vector $v^a$. This allows to convert e.g. the Dirac Lagrangian to an explicitly $SO(1,4)$ or $SO(2,3)$ invariant form by replacing all occurrences of $\theta^I$ with $\nabla v^a$. 

\subsubsection{Pure $SO(1,4)$ or $SO(2,3)$ connection formulation}

The idea of this formulation is to integrate out the vector field $v^a$ of the Stelle-West formulation. The corresponding Lagrangian has been described in \cite{West:1978nd}. Similar procedure has been considered in \cite{Freidel:2005ak} in a related context, but with the curvature squared action (\ref{action-SW}) replaced by a BF-type action containing an additional auxiliary 2-form field $B^{ab}$.

Let us add to (\ref{action-SW}) a Lagrange multiplier term to enforce the constraint. For definiteness, we consider the case of positive $\Lambda$ so that the relevant constraint is $|v|^2=1$. The action is
\be\label{action-MM-mu}
S[A,v,\mu]=- \frac{3}{64\pi G\Lambda} \int \epsilon_{abcde} {\cal F}^{ab}(A) {\cal F}^{cd}(A) v^e -\frac{\mu}{2} (|v|^2-1).
\ee
Varying this action with respect to $v$ gives
\be
\frac{1}{4} \tilde{\epsilon}^{\mu\nu\rho\sigma} \epsilon_{abcde} {\cal F}^{ab}_{\mu\nu} {\cal F}^{cd}_{\rho\sigma} \equiv  \tilde{X}_a = \tilde{\mu} v_a,
\ee
where we introduced a convenient notation, and $\tilde{\mu}\, d^4x = \mu$. The Lagrange multiplier can now be solved from the constraint and reads
\be
\tilde{\mu} = \sqrt{ |\tilde{X}|^2 }.
\ee
The resulting pure connection action \cite{West:1978nd} is the integral of the Lagrange multiplier
\be\label{action-West}
S[A]=- \frac{3}{64\pi G\Lambda} \int\sqrt{ |\tilde{X}|^2 }.
\ee
This action, however, is not very useful for a perturbative expansion. Indeed, one typically wants to expand around a maximally symmetric background which in this case corresponds to ${\cal F}^{ab}=0$. We cannot expand the square root around zero, and so (\ref{action-West}) is not useful as a starting point for gravitational perturbation theory. But the action (\ref{action-MM-mu}) one step before the pure connection action, and especially its MacDowell-Mansouri version (\ref{action-MM}) in which the de Sitter symmetry is explicitly broken to Lorentz is very convenient for developing  perturbation theory, as we saw above. 

\subsection{BF formulations}

The idea of BF-type formulations is to replace the wedge product $\epsilon_{IJKL} \theta^K\wedge \theta^L$ of two tetrads in the Einstein-Cartan action with a new 2-form field $B_{IJ}$. However, in 4D not every 2-form field $B^{IJ}$ is of the required form and one adds a set of constraints on the 2-form field to guarantee that it ``comes from a tetrad''. In 4D this has been first considered by Freidel and De Pietri in \cite{DePietri:1998hnx}, and so we will refer to the corresponding model by the initials of these authors.\footnote{Plebanski  \cite{Plebanski:1977zz} has considered essentially the same model before, as his paper also contains an action that includes both the self-dual and anti-self-dual sectors.} The higher dimensional version has been developed in \cite{Freidel:1999rr}. 

Consider the following action
\be\label{action-FdP}
S_{\rm FdP}[B,\omega,\Psi] = \frac{1}{16\pi G} \int B_{IJ}\wedge F^{IJ}(\omega) - \frac{1}{2} \left( \Psi^{IJKL} +\frac{\Lambda}{6} \epsilon^{IJKL} \right) B_{IJ}\wedge B_{KL}.
\ee
The Lagrange multiplier field $\Psi^{IJKL}$ is required to be tracefree $\Psi^{IJKL}\epsilon_{IJKL}=0$. When $B_{IJ}=(1/2)\epsilon_{IJKL}\theta^K\wedge \theta^L$ the above action reduces to (\ref{action-EC}). 

Varying (\ref{action-FdP}) with respect to the Lagrange multiplier field $\Psi^{IJKL}$ we get the constraint
\be\label{so4-constr}
B^{[IJ}\wedge B^{KL]} \sim \epsilon^{IJKL}.
\ee
As is shown in \cite{Freidel:1999rr}, Theorem 1, this equation implies that $B^{IJ}$ is either the wedge product of two frame fields, or the dual of such a wedge product
\be\label{simpl-sols}
B^{IJ} = \pm \theta_I \wedge \theta_J \qquad \text{or}\qquad B^{IJ}=\pm\frac{1}{2}\epsilon_{IJKL} \theta^K\wedge \theta^L.
\ee
The second set of solutions to the constraints (\ref{so4-constr}) is what gives GR, because the action then reduces to (\ref{action-EC}). The first set of solutions gives the so-called Holst term \cite{Holst:1995pc}. After integrating out the spin connection it becomes a total derivative. 

The Lorentz group $SO(1,3)$, in whose Lie algebra the 2-forms fields $B^{IJ}$ are valued, is not simple. The general invariant metric on the Lie algebra is an arbitrary linear combination of two metrics $\delta^{[I}_K \delta^{J]}_L$ and $\epsilon^{IJ}{}_{KL}$. In (\ref{so4-constr}) we have imposed the tracelessness of $\Psi^{IJKL}$ with respect to a particular metric from this class. It is also possible to consider a more general tracefree constraint, as was first studied in \cite{Capovilla:2001zi}. This removes the degeneracy present in (\ref{simpl-sols}) and gives a single solution, which is a linear combination of the two solutions in (\ref{simpl-sols}). The action evaluated on the solution is then the Einstein-Cartan action with the addition of the Holst term  \cite{Holst:1995pc}.

Thus, classically, the theory (\ref{action-FdP}), or its version \cite{Capovilla:2001zi} where one imposes a more general tracefree condition on $\Psi^{IJKL}$, describes GR in the sense that all solutions of GR are also solutions of this theory.  

The formulation (\ref{action-FdP}) is the starting point of the so-called spin foam model quantisation of gravity \cite{Perez:2012wv}.

\subsection{Plebanski and related formulations}

We now come to what is possibly the least familiar formulation of all. It was first introduced in a paper by Plebanski \cite{Plebanski:1977zz} 
and was later rediscovered in \cite{Samuel:1987td,Capovilla:1991qb}, in the authors' search for a Lagrangian formulation for Ashtekar's new Hamiltonian formulation of GR \cite{Ashtekar:1987gu}. A review of the Plebanski formulation is given in \cite{Krasnov:2009pu}.

\subsubsection{Decomposition of the Riemann curvature}

To motivate the Plebanski formulation we need to review some properties of the curvature specific to four dimensions. The special property of 4D is that the Hodge star maps 2-forms into 2-forms, and introduces the decomposition of the space of 2-forms into self- and anti-self-dual parts
\be\label{lambda2-split}
\Lambda^2 = \Lambda^+ \oplus \Lambda^-.
\ee
The Riemann curvature can then be viewed as a symmetric $\Lambda^2 \otimes \Lambda^2$ valued matrix. Decomposing this matrix into its $\Lambda^\pm$ components we get the following block form
\be\label{riemann}
\text{Riemann} = \left( \begin{array}{cc} A & B \\ B^T & C \end{array} \right).
\ee
Here $A,C$ are symmetric, and $B$ is an arbitrary $3\times 3$ matrix. There are also some reality properties that are signature dependent. In the Euclidean and split $(-,-,+,+)$ signature the decomposition (\ref{lambda2-split}) works with real coefficients. In the Lorentzian signature one must complexify the space of 2-forms to perform (\ref{lambda2-split}). In the case of Euclidean and split signatures all matrices $A,B,C$ are real. For Lorentzian signature the matrices $A,C$ are complex and complex conjugates of each other $C^* = A$, and $B$ is Hermitian $(B^T)^*=B$. In all cases the traces of $A,C$ are equal, and equal to the scalar curvature ${\rm Tr}(A)={\rm Tr}(C) = R/4$. One can also show that the tracefree parts of $A,C$ encode the self- and anti-self-dual parts of the Weyl curvature, while $B$ is the trace-free part of Ricci curvature. 

The observation that makes the Plebanski formulation work is that it is sufficient to have access to just one of the rows of the matrix (\ref{riemann}) to impose the Einstein condition. Indeed, the Einstein condition $R_{\mu\nu}=\Lambda g_{\mu\nu}$ can be stated as the condition that the Ricci tensor has only the trace part. In view of what was said above, this is equivalent to imposing the condition $B=0$. This can be imposed by taking the first row of the matrix (\ref{riemann}), which has the interpretation of the curvature of the self-dual part of the spin connection. Thus, we decompose the spin connection as
\be
\omega^{IJ} = A_+^{IJ} + A_-^{IJ},
\ee
where the dual is taken with respect to the internal indices. 
The curvature of $\omega^{IJ}$, which coincides with the Riemann tensor when $\omega$ is torsion free, decomposes as the sum
\be
F^{IJ}(\omega) = F^{IJ}_+ + F^{IJ}_-,
\ee
where $F^{IJ}_\pm$ are the curvatures of $A^{IJ}_\pm$. This happens because in the complex domain the Lie algebra of the Lorentz group splits as the direct sum of two ${\mathfrak su}(2)$ Lie algebras. Thus, each of the two connections $A_\pm$ is actually an $SU(2)$ connection. The decomposition (\ref{riemann}) tells us that the Einstein condition can be encoded as the statement that the curvature of the self-dual part of the spin connection is self-dual as a 2-form. The Plebanski formulation and its variants are based on this way of expressing the Einstein condition. It is clear that all this is specific to four dimensions.

\subsubsection{Chiral first order formulation}

The discussion above tells us that to impose the Einstein condition it is enough to have access to just a half of the spin connection $\omega^{IJ}$. We can take this to be the self-dual half $A_+$, which we shall from now on denote simply by $A$. To write an action that realises this idea, we recall the fact that one can add to the Einstein-Cartan action the Holst term $\theta_I \wedge \theta_J F^{IJ}$ with an arbitrary coefficient, without changing the dynamics of the theory. Indeed, when the connection has zero torsion this term becomes a total derivative. This can be easily seen by considering the torsion squared $\nabla \theta^I \wedge \nabla \theta_I$. Integrating by parts here one gets a multiple of the Holst term. 

So, we add to the Einstein-Cartan Lagrangian the Holst term with a coefficient chosen so that the self-dual part of $\theta^I\wedge \theta^J$ is taken:
\be
S_{\rm chiral}[\theta,w] = \frac{1}{8\pi G\sqrt{\sigma} } \int \theta_I\wedge \theta_J P_+^{IJ}{}_{KL} \wedge \left( F^{KL} - \frac{\Lambda}{6} \theta^K\wedge \theta^L\right).
\ee
where
\be
P_+^{IJ}{}_{KL}=\frac{1}{2} \left( \delta^{[I}_K \delta^{J]}_L + \frac{\sqrt{\sigma}}{2} \epsilon^{IJ}{}_{KL} \right) 
\ee
is the self-dual projector. Here $\sigma$ is the signature related sign, with $\sigma=-1$ for the Lorentzian signature. Thus, in the Lorentzian signature we have added to the Lagrangian the Holst term with an imaginary coefficient, and the Lagrangian is no longer manifestly real. Working with complex-valued fields, will be economic, as we shall see below, but will also lead to some headaches related to reality conditions.

The next step is to recall that the self-dual projector applied to the curvature gives the curvature of the self-dual part of the spin connection. So, we can alternatively write the above Lagrangian as
\be
S_{\rm chiral}[\theta, A] = \frac{1}{8\pi G\sqrt{\sigma} } \int (\theta_I\wedge \theta_J)_+ \wedge \left( F^{IJ}(A) - \frac{\Lambda}{6} (\theta^I\wedge \theta^J)_+ \right),
\ee
where the plus subscript on the wedge product of two tetrads could be omitted because the projection is taken automatically by contracting with the self-dual $F^{IJ}(A)$. 

This Lagrangian is written most economically in spinor notations. We remind the reader that in four dimensions there are Weyl spinors of two different types, and the tangent bundle splits as the product of spinor bundles $TM = S_+\otimes S_-$. Similarly, the bundle of 2-forms splits as $\Lambda^2=S_+^2 \oplus S_-^2$, where $S_\pm^2$ denotes the space of symmetric rank 2 spinors of the corresponding type. The self-dual connection $A$ then becomes an object $A^{AB}$, where $A,B=1,2$ are the unprimed spinor indices denoting objects in $S_+$. The tetrad is an object $\theta^{AA'}$, and the self-dual part of the wedge product of two tetrads is selected by contracting the primed spinor indices. All in all, we get the following Lagrangian
\be\label{action-chiral}
S_{\rm chiral}[\theta, A] = \frac{1}{8\pi G\sqrt{\sigma} } \int \theta_{AA'} \wedge \theta_B{}^{A'} \wedge \left( F^{AB}(A) - \frac{\Lambda}{6} \theta^{A}{}_{B'} \wedge \theta^{BB'} \right),
\ee
where the curvature is given by $F^{AB}=dA^{AB} + A^{AE}\wedge A_E{}^B$. 

The main outcome of all these manipulations is that we halved the number of the connection components that enter the Lagrangian. Indeed, in the Einstein-Cartan case (\ref{action-EC}), the Lagrangian depends on $24$ connection components per spacetime point. This is better than the case of Palatini theory (\ref{action-palatini}), where in addition to the $10$ metric components there are also $40$ components of the affine connection. But this is nevertheless quite many components to carry around in explicit calculations. What was achieved by passing to (\ref{action-chiral}) is that now, in addition to the $16$ components in the tetrad, the Lagrangian depends on just $12$ connection components. One could object that the connection is now complex, and so its real and imaginary parts continue to comprise the same $24$ components. But this is not the right interpretation. The Lagrangian depends on the $12$ components of the self-dual connection $A^{AB}$ holomorphically, as no complex conjugate connection ever appears. Also, in Euclidean signature no complexification has happened, and we indeed just halved the number of the connection components with the self-dual projection trick. 

To summarise, the ``chiral'' formulation (\ref{action-chiral}) keeps the main advantage of the Einstein-Cartan formulation of GR --- it is polynomial in the fields, with at most quartic terms appearing in the action. And it is also much more economical than the Einstein-Cartan formulation, because it depends only on $16+12$ field components per spacetime point, as compared to $16+24$ components in the Einstein-Cartan case. This makes (\ref{action-chiral}) much better suited for explicit e.g. perturbative calculations. One complication is that one needs to deal with the issue of reality conditions in the Lorentzian case. However, at least for perturbative calculations, these are not difficult to impose. One just imposes the condition that the tetrad is real, i.e. Hermitian. The correct reality conditions on the connection are then imposed automatically by the field equations. Further, loop calculations are customarily performed in Euclidean signature, and then one does not need to worry about reality conditions at all as all fields are real. 

The final remark is that, unlike in the full Einstein-Cartan formulation, in the chiral theory (\ref{action-chiral}) the Hamiltonian analysis does not lead to any second-class constraints. This is directly linked to the halving of the number of ``momentum'' variables introduced in this first-order theory. The Hamiltonian analysis of (\ref{action-chiral}) directly leads to Ashtekar's new Hamiltonian formulation of GR \cite{Ashtekar:1987gu}.

\subsubsection{Plebanski formulation}

Plebanski's formulation \cite{Plebanski:1977zz} takes one further step, and replaces the self-dual 2-form $\theta_{AA'} \wedge \theta_B{}^{A'}$ with a new 2-form field $B_{AB}$. It then adds to the action a Lagrange multiplier term that guarantees that $B_{AB}$ is a wedge product of two tetrads. This is similar to what was done in the passage from 
the Einstein-Cartan formulation to the BF action (\ref{action-FdP}). 

In spinor notations, the Plebanski action reads
\be
S_{\rm Pleb}[B,A,\Psi] = \frac{1}{8\pi G\sqrt{\sigma} } \int B_{AB} \wedge F^{AB} - \frac{1}{2} \left( \Psi^{AB}{}_{CD} + \frac{\Lambda}{3} \epsilon^{(A}{}_{C} \epsilon^{B)}{}_D \right) B_{AB}\wedge B^{CD}.
\ee
Here $\epsilon^{AB}$ is the spinor metric and the Lagrange multiplier field is required to be completely symmetric. However, given that there are now no primed spinor indices in sight, it is convenient to rewrite the Plebanski Lagrangian in $SO(3)$ notations. Thus, we replace a symmetric pair $AB$ with an index $i=1,2,3$. The connection is then an $SO(3)$ (complexified, in the case of Lorentzian signature) connection. The action reads
\be\label{action-Pleb}
S_{\rm Pleb}[B,A,\Psi] = \frac{1}{8\pi G\sqrt{\sigma} } \int B^i \wedge F^i - \frac{1}{2} \left( \Psi^{ij} - \frac{\Lambda}{3} \delta^{ij} \right) B^i\wedge B^j.
\ee
Varying this action with respect to the Lagrange multiplier field $\Psi^{ij}$, which in the $SO(3)$ notations is required to be tracefree, we get the constraint 
\be\label{so3-constr}
B^i \wedge B^j \sim \delta^{ij},
\ee
which can be compared to (\ref{so4-constr}). This constraint implies that $B^i$ can be written as (plus or minus) the self-dual part of the wedge product of two tetrads. We are then back to the chiral formulation (\ref{action-chiral}), and so we get a formulation of GR with ${\mathfrak so}(3)$ valued 2-form field $B$ and connection $A$ as the basic variables. 

Now that the basic variable is a 2-form field, it is not clear how to obtain the metric. As we have said, 
when $B^i$ satisfies (\ref{so3-constr}),
there exists a tetrad that gives this 2-form field. 
However, it would be more convenient to have an explicit formula for the metric in terms of $B^i$. 
Such a formula exists and is known in the literature as the Urbantke formula \cite{Urb}. It gives a densitized metric
\be\label{metric-Urb}
\tilde{g}_{\mu\nu} = \frac{1}{12} \tilde{\epsilon}^{\alpha\beta\gamma\delta} \epsilon^{ijk} B^i_{\mu\alpha} B^j_{\nu\beta} B^k_{\gamma\delta}.
\ee
The metric itself can be computed by noting that the volume form is given by the sixth root of the determinant of the right-hand-side. 

Apart from the constraint (\ref{so3-constr}), the other field equations that follow from (\ref{action-Pleb}) are as follows
\be\label{pleb-feqs}
d_A B^i =0, \qquad F^i = \left( \Psi^{ij} + \frac{\Lambda}{3} \delta^{ij} \right) B^j.
\ee
The first of these equations is the analog of the torsion-free condition in the Plebanski setup. Together with the constrain (\ref{so3-constr}) it implies that $A$ is the self-dual part of the spin connection compatible with the metric (\ref{metric-Urb}). The second equation then states that the curvature of the self-dual part of the spin connection is self-dual as a 2-form, which we know to be equivalent to the Einstein condition. As we know from (\ref{riemann}), the self-dual-self-dual block $A$ of the Riemann curvature tensor is just the self-dual part of the Weyl curvature plus a multiple of the scalar curvature. So, the second equation in (\ref{pleb-feqs}) also says that on-shell the Lagrange multiplier field $\Psi^{ij}$ receives the interpretation of the self-dual part of the Weyl curvature. 

In Lorentzian signature all fields are complex-valued, and so one must impose appropriate reality conditions. As in the chiral first order formulation described above, it is sufficient to impose the reality conditions on the metric-like field $B^i$, the appropriate reality condition on the connection then gets imposed automatically by the field equations. The conditions on the 2-form field are
\be\label{reality}
B^i \wedge (B^j)^* =0, \qquad {\rm Re}\left( B^i \wedge B^j\right) =0.
\ee
The first of these equations gives 9 conditions which guarantee that conformal class of the metric (\ref{metric-Urb})  is real, while the last condition gives the reality of the volume form. 

We remark that the Plebanski formulation, as well as the related formulation (\ref{action-FdP}), is cubic in the fields, even with non-zero cosmological constant. This is the only known formulation of GR with $\Lambda\not=0$ that is cubic. However, a drawback of this formulation is that it is not so easy to couple spinors to two-forms. The only known way of doing this is described in \cite{Capovilla:1991qb} and uses further Lagrange multipliers. 

\subsubsection{Chiral pure connection formulation}

The 2-form field of the Plebanski formulation can be integrated out, resulting in the action
\be\label{action-a-psi}
S[A,\Psi] = \frac{1}{16\pi G\sqrt{\sigma} }\int \left( \Psi^{ij} - \frac{\Lambda}{3} \delta^{ij} \right)^{-1} F^i \wedge F^j.
\ee
This action, which is an intermediate step towards the pure connection formulation below, is itself a useful variational principle for GR. It depends on just $12+5$ variables. Even though it appears to be second-order in derivatives, this is an illusion. The most natural backgrounds on which this action can be expanded are maximally symmetric. On such backgrounds $\Psi^{ij}=0$ (zero Weyl curvature), and the part of the linearised action that is quadratic in derivatives is just $d_A \delta A^i \wedge d_A \delta A^i$. Integrating by parts and replacing the commutator of covariant derivatives with a curvature one reduces this to a term not containing derivatives.

The action (\ref{action-a-psi}) exists even with $\Lambda=0$, but in this case it is not possible to expand it around a $\Psi^{ij}=0$ background. This action is surprisingly similar to the MacDowell-Mansouri action (\ref{action-MM}) in that it is obtained as the wedge product of two copies of the curvature, contracted with some appropriate tensor. The similarity becomes even more pronounced if one compares to the action (\ref{action-SW}) that contains a dynamical field in front of the curvature squared term. 

To go to the pure connection formulation we do the trick that we already applied several times --- we add to the action a Lagrange multiplier term imposing the relevant constraint on the field that appears in front of the wedge product of curvatures. We have already used this trick in passing to the pure connection formulation related to MacDowell-Mansouri. 

Thus, let us write the action (\ref{action-a-psi}) as
\be
S[A,\Psi,\mu] = \frac{1}{16\pi G\sqrt{\sigma} }\int \left( M^{ij} \right)^{-1} F^i \wedge F^j + \mu \left({\rm Tr}(M) - \Lambda\right).
\ee
Note the perfect similarity between this action and (\ref{action-MM-mu}). We now integrate out $M^{ij}$. Its Euler-Lagrange equation reads
\be
(M^{-1})^{ik} F^i \wedge F^j (M^{-1})^{lj} = \mu \delta^{ij},
\ee
and so if we introduce
\be
\tilde{X}^{ij} := \frac{1}{4} \tilde{\epsilon}^{\mu\nu\rho\sigma} F^i_{\mu\nu} F^j_{\rho\sigma},
\ee
and write $\mu = \tilde{\mu} \, d^4x$ we get
\be
M^{ij} = \left(\sqrt{\frac{\tilde{X}}{\tilde{\mu}}}\right)^{ij}.
\ee
As usual, the Lagrange multiplier $\tilde{\mu}$ is found from the constraint it imposes, and the pure connection action becomes the integral of the Lagrange multiplier
\be\label{action-K}
S_{\rm K} = \frac{1}{16\pi G\Lambda \sqrt{\sigma} }\int \left( {\rm Tr}\sqrt{\tilde{X}} \right)^2.
\ee
This action was first obtained in \cite{Krasnov:2011pp}. It is the most economic pure connection formulation of GR available. Indeed, it must be compared to the action (\ref{action-West}) that depends on the $4\times 10$ components of the connection, and to the linearised action (\ref{pure-spin-action-2}) that depends on the $24$ components. In contrast, (\ref{action-K}) depends on just $12$ components of the $SO(3)$ connection. It is thus comparable to the usual metric formulation with its 10 components in economy. Moreover, it turns out that the perturbation theory in this chiral pure connection formalism can be set up in such a way that only $8$ out of the $12$ components propagate, 2 of them being the physical polarisations of the graviton, the remaining $3+3$ being unphysical gauge variables, see \cite{Krasnov:2011up}. This is more economical than GR in the metric formalism, where, having fixed a gauge, $10$ components propagate, 2 of them being the physical polarisations of the graviton. But this perturbation theory only exists around $\Lambda\not=0$ backgrounds, because of the presence of $1/\Lambda$ in front of the action. 

\subsection{Summary}

We now summarise the above constructions. 
We can divide the formulations of GR into two classes, depending on the group of gauge transformations that leaves the Lagrangian invariant. 
One class consists of the metric and related formulations. The gauge group of these formulations is $\Diff M$, the diffeomorphisms of spacetime. No ``internal space'' is introduced in these formulations: they work with spacetime and its tangent bundle. So, even if one introduces an independent connection to obtain a first-order formalism (\ref{action-palatini}), this is a connection in the tangent bundle.  

All other formulations can be interpreted in terms of a bundle $E$ over spacetime with fibers being copies of some internal space.\footnote{As mentioned in section III.B, this interpretation is not strictly necessary, but it strongly motivates the approach to unification that we shall discuss later.}
There is then a connection acting on sections of this bundle. The field encoding dynamical information is the soldering form, or a component of the connection as in MacDowell-Mansouri formulation. The group of local gauge symmetries in all formulations of this type is the (semi-direct) product of $\Diff M$ with a group of local gauge transformations of the fibers. In some of these formulations the basic dynamical fields are differential forms, and the Lagrangian is constructed as the wedge product of forms. These formulations are particularly attractive, because
they are polynomial.

We did not discuss in detail the coupling of gravity to other fields. Given that the philosophy is to get (most optimistically all) the bosonic fields by enlarging the gravitational gauge group, we do not need to discuss this. However, fermions will never arise from bosonic constructions of the type envisaged. So, they have to added  by hand. How to do this depends on the specific scenario. 

\subsection{Linear vs. non-linear realizations}

In all known formulations of GR, the theory is power-counting non-renormalisable. 
Furthermore, the dynamical field encoding information about the metric is always non-linear,
due to the constraints on the signature of the metric,
and the nondegeneracy of the soldering form.
In the world of flat space QFTs there is a class of non-renormalisable models that exhibit very similar features: the non-linear sigma models. 

For example, let us consider the chiral models, which are particular non-linear sigma models with values in a Lie group. 
These have actions of the form
\be\label{chiral-model}
S=-\frac{1}{2}f_\pi^2\int d^4x\,\tr(U^{-1}\partial U)^2
\ee
where $f_\pi$ has dimensions of mass.
To exhibit the analogy between these models and gravity, we note that, by discarding a total derivative term, the Hilbert action can be written in the schematic form
\be
S=m_P\int d^4x\sqrt{-g}\,\Gamma\Gamma
\ee
where $\Gamma$ are the Christoffel symbols, see e.g. \cite{Landau:1982dva}, Chapter 93.
These have the structure $\Gamma=g^{-1}\partial g$, so that the gravitational action looks very similar to the chiral action. Both actions are non-polynomial (when expanded around a background the action contains infinitely many vertices), have a dimensionful 
coupling and are power-counting non-renormalizable.

The non-linear sigma models 
can be constructed from free scalar field theory by adding a set of constraints. For example, the simplest non-linear sigma model is obtained from a set of scalars taking values in $\R^n$ by 
imposing the condition that the scalars take value in 
the sphere $S^{n-1}\subset \R^n$. In the case of $S^3=SU(2)$ we get the chiral model (\ref{chiral-model}). 

The non-renormalisable sigma model becomes renormalisable if one replaces the $\delta$-function type constraint with a quartic potential designed so that the minimum of the potential corresponds to the required submanifold. This adds to the theory an extra propagating degree of freedom, which in the SM is the Higgs field. It is thus very tempting to think that the same mechanism may also be at work in gravity, and that the non-renormalisability can be cured by replacing non-linear fields with linear ones.

A remark is in order about the tetrad and BF-type formulations. The corresponding Lagrangians are written in terms of differential forms and are polynomial, unlike the Lagrangian in metric formulation. Differential forms can be added, and so it may seem that we have a linear realisation here. However, if we try to expand the Lagrangian written in terms of tetrads around the zero configuration, there is no quadratic term, so no useful perturbation theory arises. And if we rewrite the theory in BF form, where one can now expand the kinetic $B\wedge F$ term around the zero configuration, the constraints present in the potential-type terms prevent us from getting a useful perturbative expansion around the zero vacuum. This is most clearly seen in the formulations that are intermediate steps before the pure connection formulation, see e.g. (\ref{action-a-psi}).  These are of the Stelle-West type (\ref{action-MM-mu}) and contain a non-linear constraint on the auxiliary field. 

The situation is slightly different for the MacDowell-Mansouri formulation. Here the field is a De Sitter (or anti De Sitter) connection. The vacuum corresponds to a flat connection. So, it could be taken to be the zero connection. However, given that the metric is a part of the connection, one needs to explain why a particular flat connection that gives a non-zero metric is selected. The non-degeneracy of the metric is thus not automatic in this formalism. This is similar to all other formalisms where the non-degeneracy of the metric field is part of the definition of the theory. 
Furthermore, in the Stelle-West formulation
the field $v^a$ is subject to a non-linear constraint
that is very similar to that of a spherical non-linear sigma model.
Thus, we conclude that none of the discussed formulations of General Relativity is in terms of linearly realised fields, even when differential forms are used.

This discussion suggests that the non-renormalisability of gravity and the non-linear nature of its basic field (in particular its non-zero VEV) are related, and that the non-renormalisability may be cured by adding extra degrees of freedom (Higgs fields) so as to convert a non-linear realisation (group manifold or a group coset) into a linear one (vector space). 
However, nobody has been able to realise these ideas.
One important difference is that relaxing the constraints
in GR should presumably not introduce new degrees of freedom,
because they are in the form of inequalities (``anholonomic'') rather than equalities (``holonomic'').

The only situation where the idea of linear realisation works is 3D gravity \cite{Witten:1988hc}. 
In this case the Einstein-Cartan action is cubic in the fields, and has a perturbative expansion around the zero frame field configuration. This is related to the fact that the MacDowell-Mansouri type of formulation of 3D gravity is just the Chern-Simons theory of the corresponding De Sitter or anti De Sitter connections. The space of connections is diffeomorphic to a linear space, and so we have essentially a linear realisation that moreover admits a good perturbative expansion around the zero field configuration. 

Thus, in spite of this idea being attractive, whether gravity can be described in terms of linearly realised fields, and whether this can cure its non-renormalisablity remains open. 
We will not make any new proposal along these lines here.

\section{Hints of unification}
\label{sec:hints}

Before studying in more detail some models that unify gravity with the bosonic fields of the type present in the SM, let us ask whether there is any evidence for this kind of unification in the real world. As with all other extensions of the SM and GR, one can give only rather weak circumstantial evidence, but it is worth pointing it out at once. 

\subsection{Convergence of the couplings}

A crucial aspect of a unified theory, as spelled out in section 
\ref{sec:whatis}, is that the coupling constant at high energy is unique.
Below the unification scale, the gauge couplings relative to different 
gauge groups run differently and are not expected to be equal.
One of the main arguments in favor of GUT theories is the fact that the 
gauge couplings
$\alpha_1$, $\alpha_2$ and $\alpha_3$ of
the groups $U(1)_Y$, $SU(2)$ and $SU(3)$ tend to converge as the energy 
increases.
If nothing more than the SM existed, the renormalization group 
trajectories would not cross at a single point. This has been used for a 
long time as an argument in favor of supersymmetry.
However, there could be many other intermediate states beyond the 
present reach of accelerators that could change the beta functions and 
make the three trajectories cross at a single point.

How does gravity fit in this picture?
Unlike the couplings of the SM, the gravitational coupling is 
dimensionful. We can form a dimensionless coupling $\tilde G$, analogous 
to $\alpha_1$, $\alpha_2$ and $\alpha_3$, multiplying Newtons' constant 
by the square of an energy.
In a collision process, this could be one of the Mandel'stam variables, 
for example.
This coupling $\tilde G$ has the property that it depends on the energy 
already at the classical level. It has a classical beta function 
$2\tilde G$.
Due to the fact that the energies we can reach are so much smaller than the
Planck energy, $\tilde G$ is very small, of the order of $10^{-16}$ for 
particles at the LHC.
This is why gravity is negligible in particle physics. On the other 
hand, $\tilde G$ runs much faster than the other couplings: 
quadratically instead of logarithmically.
Thus $\tilde G$ becomes of the same order as the other couplings at the 
Planck scale.
It is remarkable that in many GUTs, the energy scale at which the 
crossing, or near-crossing, happens is only a few orders of magnitude 
below the Planck scale.
One can take this as a hint in favor of a unification that also involves gravity \cite{wilczek}.

\subsection{Kaluza-Klein hint}

The bosonic fields that we know to exist and appear in the SM coupled to gravity are: (i) the metric to describe gravity; (ii) gauge fields charged with respect to the SM gauge group $SU(3)\times SU(2)\times U(1)$; (iii) the Higgs field. 
Other fields, whose existence has not yet been verified, may be needed for specific models, for examples an inflaton \footnote{Unless the Higgs field is used for this purpose as in \cite{Bezrukov:2007ep}}, as well as fields to describe dark matter. However, since we don't want to have several spin 2 fields around, it is not very restrictive to assume that any such bosonic fields will again be either scalars or gauge fields. A very compelling scheme where all such fields can be described as components of a single field is Kaluza-Klein (KK) theory, where they are all interpreted as components of a higher-dimensional metric. Schematically,
\be\label{KK}
\text{METRIC} = \left( \begin{array}{cc} \text{Higgs} & \text{Connection} \\ \text{Connection} & \text{Metric} \end{array} \right)\ .
\ee
In spite of difficulties with dynamical realisations of this idea, see section \ref{sec:KK}, it still remains one of the strongest hints that gravity should be unified with the other known bosonic fields.

\subsection{Fermions}
\label{sec:fermions}

The orthogonal groups $SO(2k)$ and $SO(2k+1)$ have spinor representations. These can be given a simple geometrical construction.
Consider first the complexified groups, in order to avoid having do deal with different possible signatures. 
The spinor representation of $SO_\C(2k)$, can be constructed as the space of all differential forms in 
${\mathbb C}^k$. It has dimension $2^k$ and is reducible. The irreducible subspaces consist of even and odd degree differential forms, each of dimension $2^{k-1}$. The spinors taking value in these irreducible representations are called Weyl spinors. In the setting over reals the structure of spinor representations depends on the dimension as well as the signature. The possibilities are complex, real and quaternionic spinor representations. A useful source for this material is \cite{Jose}.

As is well-known, see e.g. \cite{Baez:2009dj} for a nice description, all fermions of the single generation of the SM, 
supplemented with the right-handed neutrino that is required to explain the neutrino oscillations, fit into the single $16$-dimensional (complex) Weyl representation of the group $SO(10)$. 
To see this, it is clearest to count using the 2-component spinor formalism, as is reviewed in e.g. \cite{Dreiner:2008tw}. Then each SM fermion is described using two unprimed (left-handed)
2-component Lorentz spinors. 
(The right-handed components of each particle are described
as the charge conjugate of a left-handed spinor).
The only particle requiring a single unprimed 2-component spinor is the left-handed neutrino. But one usually extends the SM adding the right-handed neutrino. 
Then the 2-component Lorentz spinor content of one SM family is:
a weak doublet consisting of left-handed neutrino and electron, as well as 3 doublets for 3 colours of the left-handed up and down quarks. This gives in total 8 2-components spinors. Plus there is the same number of unprimed 2-component spinors that are all weak singlets.
This gives 16 2-component unprimed spinors. These form the 16-dimensional Weyl representation of $SO(10)$. 

The SM gauge group $SU(2)\times U(1)\times SU(3)$, modulo a certain discrete subgroup, see e.g. \cite{Baez:2009dj}, can be embedded first into $SU(5)$, which in turn is a subgroup of $SO(10)$. In the realisation of the Weyl representation as differential forms in $\R^5$, the subgroup $SU(5)$ mixes the forms of a fixed degree, without changing the degree of the form. Thus, if we realise the Weyl representation in question by, say, even forms, the 16-dimensional Weyl representation splits as the 1-dimensional space of 0-forms, plus 10-dimensional space of 2-forms, plus 5-dimensional space of 4-forms. These are all irreducible representations of $SU(5)$. The 1-dimensional representation describes the right-handed neutrino, the 5-dimensional representation describes the 3 colours of the right-handed down quark plus the left-handed electron-neutrino doublet, and the 10-dimensional representation describes the colour triplet and weak doublet of the left-handed up and down quark, plus the colour triplet of the right-handed up quark, plus the right-handed electron. For more details on this standard GUT material see \cite{Baez:2009dj} for a somewhat more mathematically oriented exposition, and e.g. \cite{Srednicki:2007qs}, Chapter 97 for textbook treatment. 

All the described fermions are also spinors of the 4-dimensional Lorentz group $SO(1,3)\sim SL_\C(2)$, but the Lorentz group did not play any role in the above discussion.
Now the spinor representations of $SO(2k)$ have the property that if one takes $SO(2p),$ $p<k$ and embeds it into $SO(2k)$ 
in the obvious way, so that the commutant of this embedded $SO(2p)$ is $SO(2(k-p))$, the Weyl spinor of $SO(2k)$ splits as a direct sum of Weyl bi-spinors, i.e. spinors of $SO(2p)$ as well as spinors of $SO(2(k-p))$. Thus, spinors of bigger orthogonal groups decompose as spinors of their smaller orthogonal subgroups. This follows quite directly from the differential forms construction of the spinor representations. 

We can attempt to use this fact to embed the $SO(10)$ GUT gauge group together with the Lorenz group $SO(4)$ into $SO(14)$ \cite{percacci1,percacci2}. Again, at first everything is viewed over complex numbers to avoid having to deal with different possible signatures. 
Then the Weyl representation of $SO(14)$ is 64-dimensional. If we embed $SO(10)\times SO(4)$ so that they commute, the 64-dimensional representation splits as the 16-dimensional Weyl representation of $SO(10)$ which is also the unprimed 2-component spinor of $SO(4)$, plus the other 16-dimensional Weyl representation of $SO(10)$, which is the primed 2-component spinor of $SO(4)$
\be\label{spinor-splitC}
\mathbf{64}=\mathbf{2}\times\mathbf{16}\oplus \overline{\mathbf{2}}\times\overline{\mathbf{16}}
\ee
The first multiplet on the right-hand-side corresponds to the fermionic content of one SM family, now with the Lorentz group spinor indices taken into account.

Let us discuss the same picture over the reals. If we consider groups $SO(p,q)$ with $p+q=14$ and containing $SO(10)\times SO(1,3)$ as a subgroup,
there are only two possibilities: $SO(3,11)$ and $SO(1,13)$. As is well-known, see e.g. \cite{DAuria:2000byu} for a review, or \cite{Jose} for a more concise description, the type of spinors one gets for $SO(p,q)$ in the real case is governed by the signature $(p-q) {\rm mod}(8)$. Among even signatures, 
signature zero gives a real representation, signature 4 a quaternionic representation, signatures 2 and 6 give complex representations, see e.g. Table 2 in \cite{DAuria:2000byu}. 
In the case of the group $SO(3,11)$ the signature is zero, and the spinor representation is 64-real-dimensional
(it is called a Majorana-Weyl representation). 
Under the embedding $SO(1,3)\times SO(10) \subset SO(3,11)$ this real representation splits as 
\be\label{spinor-splitR}
\mathbf{64}_\R=\mathbf{2}_\C\times\mathbf{16}_\C
\ee
which is exactly what is needed for one generation of the SM
\cite{gravigut}. 

In the case $SO(1,13)$ the signature is equal to 4, which means that the spinor representation is quaternionic, of real dimension 128. 
This is twice more than is needed to describe 
the fermions of one SM family.

A potentially interesting alternative arises if instead of demanding $SO(10)$ to be the subgroup, one only requires the Pati-Salam $SO(4)\times SO(6)$ to be embeddable. This gives a twice larger list of acceptable groups, see \cite{Maraner:2003sq}. In particular, it is now possible to consider the group $SO(7,7)$ that, similarly to $SO(3,11)$ is of signature zero and thus has a real 64 dimensional Weyl representation.

To summarise, all fermions of a single generation of the SM can be viewed as forming a single irreducible spinor representation of a 
``graviGUT'' group whose complexification is $SO_\C(14)$. This suggests that the SM gauge group, or one of its GUT extensions, should be put together with the Lorentz group, which is what the unification schemes to be described below will do. 

\subsection{The low energy effective theory of gravity}

We shall now review indications that a Higgs mechanism may be taking place in gravity. Insofar as the Higgs mechanism is usually associated with unification, this may be taken as a hint for a form
of unification. 

As we already discussed above, and as has been pointed out since long and by many authors, GR has deep similarities to the chiral models of strong interactions, or more generally to nonlinear sigma models. This is in particular due to the fact that the metric tensor is in reality a very non-linear object, already at the kinematical level. Indeed, the constraints on its eigenvalues select a subspace of symmetric tensors diffeomorphic to the coset $GL(n)/O(p,q)$, where $(p,q)$ is the signature of the metric. Likewise, the tetrad has to be non-degenerate and that makes the space of tetrads diffeomorphic to the linear group.

On the other hand, the chiral models are regarded as low-energy effective field theories of the strong interactions, valid up to energy of order $\sim f_\pi$ (omitting numerical factors) \cite{gl}. 
The target space $G$ can be viewed as the coset $G_L\times G_R/G_V$, where $G_L$ and $G_R$ act on the target space from the left and from the right, respectively, and $G_V$ is the diagonal subgroup. In general, a nonlinear sigma model with target space $G/H$ is the low-energy effective theory describing a (global) symmetry $G$ that is spontaneously broken to $H$.
The coupling $f_\pi$ is related to the scale of the breaking.
From this point of view it is natural to interpret GR as a low-energy effective field theory
\cite{donoghue1,donoghue2,donoghue3,Khriplovich:2002bt,BjerrumBohr:2002kt,BjerrumBohr:2002ks,burgess}, with the Planck mass as the temperature of a phase transition, separating the low-temperature phase of ``gravity as we know it'' from a high-energy phase in which the linear group is unbroken. 

It is not very clear what kind of physics this high-energy phase would describe.
But even before coming to that, the situation in gravity is more complicated because the linear group is gauged (as discussed in section \ref{sub-sec:gln}). Therefore the phase transition must separate not a broken/unbroken phase in the ordinary sense, but rather a low-energy Higgs phase, where the gauge fields are massive (or perhaps confined, see \cite{smilga1,smilga2,holdom,donoghue4,donoghue5,donoghue6}), 
from a high-energy phase where the gauge fields are massless.
Is there any sign of the gravitational connection being massive? In GR (independently of the fields one uses to describe it) the connection is not a propagating degree of freedom. This is indeed what one would expect to see if the connection (more precisely: the difference between the dynamical connection and the Levi-Civita connection, which is a composite field of the metric) had a mass that is much larger than the presently accessible energies.
The terms quadratic in torsion (and possibly non-metricity), which are unavoidable when gravity is viewed as an effective field theory containing also an independent connection, are just a gauge-invariant way of writing a mass term for this field.
We are then led to a picture where a kind of Higgs phenomenon occurs in gravity, giving mass to the difference between the independent connection field and the Levi-Civita connection
\cite{percacci1,percacci2,Kirsch:2005st,Leclerc:2005qc,Boulanger:2006tg}.
This is a natural explanation of the fact that in GR the connection is not an independent field, a fact that otherwise is simply postulated for reasons of simplicity.
\footnote{In Palatini, and other first order formulations of gravity, the connection is forced to be the Levi-Civita connection by the equations of motion. However, this is only a property of the simplest gravitational Lagrangians: when one includes terms with curvature squared, which are unavoidable in the effective field theory, the connection becomes an independent propagating degree of freedom.
If it is massive, it disappears from dynamics at sufficiently low energy. The Palatini formulation corresponds to taking the
limit when the mass goes to infinity.}

In particle physics the Higgs phenomenon is generally used in the context of unification, as a way to generate a distinction between different types of low-energy interactions. If a Higgs phenomenon occurs in gravity, as the previous discussion suggests, then it is natural to think that it may have something to do with unification.
Following the logic of section \ref{sec:PP-unific}, one would have to find an order parameter giving rise to the distinction between gravitational and non-gravitational gauge interactions.
This is not difficult, as we shall discuss below. What turns out to be difficult is to write a dynamics that describes correctly both the low- and high-energy phases.

\section{Unified theories}
\label{sec:unific}

A unified theory of gravity must contain pure gravity, and so a possible way to obtain such a theory is to enlarge some of the structures that are present in gravity to begin with. The Kaluza-Klein approach is to extend the four-dimensional spacetime metric to a metric in a higher-dimensional space. However, there is a natural alternative.

With the exception of the metric formulation, all formulations reviewed above  contain a connection field that defines the notion of parallel transport on some ``internal'' bundle $E$ over the spacetime manifold. Correspondingly, the group of local transformations that leaves the Lagrangian invariant is the semi-direct product of the group of diffeomorphisms of the manifold with some group of ``vertical'' transformations of the fibers. The related connection is either the Lorentz connection in the tetrad and BF formulations, an $SO(1,4)$ or $SO(2,3)$ connection in MacDowell-Mansouri framework, and the self-dual part of the Lorentz connection for Plebanski-type formulations. 

A natural approach to unification is to allow the structure group of the bundle in question to become larger than required by GR. As we have already mentioned in the Introduction, this has first been suggested by Einstein and Mayer in \cite{EM1,EM2}, in the context of unification of gravity with electromagnetism. We shall briefly review this below. In such an approach the gauge field will be a matrix-valued one-form with the general structure
\be
\label{graviGUT}
\text{GAUGE FIELD} = \left( \begin{array}{cc} \text{GUT gauge field} & \text{mixed gauge field} 
\\ 
\text{mixed gauge field} & \text{gravitational connection} \end{array} \right)\ .
\ee
It is clear that this is very different from the Kaluza-Klein, or more generally higher-dimensional approaches to unification, because one is not enlarging the spacetime but rather the internal spaces of the theory. There are however relations between these approaches that we shall discuss in section \ref{sec:eggkk}.

After a brief review of the usual Kaluza-Klein approach and its modern string theory incarnations, in this section we will occupy ourselves with ``enlargement of the Lorentz group'' 4D unification scenarios. The relevant literature is much smaller, and a comprehensive review is possible. The various proposals for unified theories along the lines of ``enlarging the gauge group'' are all extensions of one of the formulations of GR discussed in section \ref{sec:formulations}, and therefore are listed in the same order.

We begin our description 
with the Einstein-Cartan formulation. Unification in this approach has been studied for longer and in more detail. This type of unification is a rather direct extension to gravity of the notion of unification as understood in particle physics, so in this case we shall try to follow in some detail the list of steps presented in section \ref{sec:PP-unific}. We will thus discuss separately the kinematical aspects, the fermionic dynamics and the bosonic dynamics. In the other cases we shall not split the discussion in the same way.

\subsection{Kaluza-Klein unification}
\label{sec:KK}

Gravity differs from all other interactions in that it describes the dynamics of the spacetime geometry. It is only to be expected, therefore, that a unified theory containing gravity should also have a strong geometrical flavor.  As mentioned before, a unified theory must extend some of the structures that are present in the original theories.  One of the earliest and most fruitful ideas is to enlarge the spacetime by introducing extra dimensions. This allows to unify spin two, spin one and spin zero fields, as is sketched in (\ref{KK}). 

As is well-known, Yang-Mills fields with gauge group $G$ can be interpreted as connections in a principal $G$-bundle. Kaluza-Klein theory is essentially the Riemannian geometry of this principal bundle, where the metric in the base space and the metric in the group, together with the assumption that vertical and horizontal spaces are orthogonal, define a metric in the principal bundle.
In the physics literature, this point of view has been originally discussed in \cite{dewitt}.

It is also important to emphasise that the dimension of the internal space does not have to be as large as the dimension of the gauge group one desires to obtain, if one compactifies on coset spaces of the type $G/H$. 
The minimal dimension of the internal space with the group of isometries equal to the SM gauge group is 7, see \cite{Witten:1981me}.
\footnote{For KK theories with coset spaces as fibers see also
\cite{Percacci:1982yi}.}
This points towards an 11D metric as an appropriate single object to put together all known bosonic fields. 
The concrete implementation of this unification program  meets
several difficulties. We will describe this only briefly, a more comprehensive review is e.g. \cite{Duff:1986hr}.

The first difficulty is that one would like the higher-dimensional
background geometry to arise as a stable solution of the field equations (this is called ``spontaneous compactification''),
but this is not so easy to achieve, as is discussed in Chapter 1 of \cite{Duff:1986hr}. 
Leaving aside torus compactifications, which only give rise to abelian
gauge groups, all dimensions except four are supposed to form a compact,
highly curved space. This requires extra fields whose energy-momentum
tensor provides the source of this curvature.
Suitable solutions have been found using nonlinear sigma models
as sources \cite{Gava:1979sq,Omero:1980vx,gellmann1,gellmann2,gellmann3}
or gauge fields \cite{RandjbarDaemi:1982rm,RandjbarDaemi:1982hi}.
However, the spectrum of excitations around these solutions
often shows instabilities \cite{RandjbarDaemi:1983bw}, and furthermore the would-be KK gauge fields
have large (typically Planckian) mass, 
thereby defeating the original purpose of these theories.
In 11-dimensional supergravity there is a differential form
that can be used to trigger compactification via
the co-called Freund-Rubin mechanism \cite{Freund:1980xh}, as is discussed in Chapters 2, 3 of \cite{Duff:1986hr}. 
As pointed out in \cite{Duff:1986hr} Chapter 13,
truncations on the spectrum of states will generally lead to
inconsistencies.

Even when a spontaneous compactification can be achieved, there is a difficulty obtaining chiral fermions, as was anticipated already in \cite{Witten:1981me}. 

The third difficulty is getting a realistic value of the cosmological constant. The Freund-Rubin solution with a positively curved internal space (as would be required to get a non-trivial group of isometries to serve as the 4D gauge group) gives the value of the 4D cosmological constant proportional to the scalar curvature of the internal space. This is way too big if one wants Planck-size internal space. 

Finally there is the obvious fact that higher-dimensional quantum field theories have worse quantum behavior than the
corresponding four-dimensional ones.
One may not worry too much about UV completions as long as
only low energies are considered, but the likely
compactification scale is expected to be comparable to
the scale where quantum effects in gravity become important.

For all these reasons, with the ``first superstring revolution''
the attention of the community shifted to higher-dimensional theories of a different type.
First, one does not try anymore to obtain the matter fields from components
of the higher-dimensional metric. Matter fields are already present in the higher-dimensional theory.
Second, one compactifies higher-dimensional supergravity on a Ricci flat compact manifold with a parallel spinor. This can be a Calabi-Yau 6D manifold if one compactifies from 10D to 4D, or a holonomy $G_2$ manifold if one goes from 11D to 4D. Such manifolds have no non-trivial isometries, and so no gauge group arises by the usual Kaluza-Klein mechanism. However, such compactifications preserve supersymmetry, and so the effective 4D cosmological constant is zero. Its non-zero observed value should then be explained by some other mechanism, but at least one is not facing the problem of Planck size cosmological constant (of negative sign) that is generated by Freund-Rubin solutions. 

Both the gauge group and chiral fermions then arise from singularities of the compact manifold, which are made sense of using string theory. We refer to \cite{Witten:2001bf} for a description of models of this sort in the context of $G_2$ compactifications of M-theory. Thus, the modern string theory unification scenarios no longer follow the geometric pattern (\ref{KK}).

\subsection{Einstein-Cartan-type unification}
\label{sec:ec}

\subsubsection{Einstein-Mayer theory}
\label{sec:em}

The Einstein-Mayer theory developed in \cite{EM1,EM2} can be viewed as a precursor to ``unification by enlarging the gauge group''. This theory has later been studied by Cartan \cite{Cartan-EM}
and a non-abelian generalization has been discussed in \cite{rosentauber}.

With the purpose of obtaining a unified theory of gravity and electromagnetism  in mind, the authors consider objects taking values in a 5-dimensional vector space $V^5$, in addition to being tensors from the point of view of spacetime, that remains four-dimensional. For consistency of notation with our exposition, we denote the indices in $V^5$ by $I,J=1,\ldots, 5$. The 5-dimensional vector space $V^5$ is assumed to be equipped with a metric $\eta^{IJ}$. The main object in \cite{EM1} is then a mixed tensor $\theta_\mu^I$, where $\mu$ is a spacetime index.
\footnote{The papers \cite{EM1,EM2} use instead latin indices
for spacetime tensors and Greek indices for $V_5$-tensors.
The object we call $\theta^I{}_\mu$ is denoted $\gamma^\iota{}_q$
and the connection is denoted $\Gamma^\iota_{\pi q}$.}
The spacetime metric is then assumed to be given by $\eta_{IJ} \theta_\mu^I \theta_\nu^J = g_{\mu\nu}$. Thus, the object $\theta$ is just an enlarged or generalised tetrad of the type we consider in more detail below.

The other main object of the formalism \cite{EM1} is the connection $\omega_\mu{}^I{}_J$. It is assumed from the outset that $\omega$ is a metric, i.e. an $SO(5)$ connection. This is the condition (I) of the paper \cite{EM1}. There are two more conditions imposed on $\omega_\mu{}^I{}_J$, whose geometrical meaning is clarified in \cite{Cartan-EM}. Their purpose is to partially fix the connection, while still leaving a part of it that can be interpreted as the electromagnetic connection free. The main difference with the schemes considered below is that the authors in \cite{EM1} do not impose the condition that the full covariant derivative of the generalised frame $\theta_\mu^I$ is zero. 

The final outcome of the paper \cite{EM1} is a unified theory of gravity and electromagnetism, where the latter does not have sources.
This was considered unsatisfactory and motivated the further developments in \cite{EM2}. We now know that all sources should come from fermions (or other electrically charged fields), and so obtaining a bosonic theory that leads to Maxwell equations in vacuum is not unsatisfactory. We will encounter another instance of such a unified theory, possibly even more elegant than Einstein and Mayer's, in Section \ref{sec:Robinson}.

\subsubsection{Kinematics}

We now consider a more general variant of this unification scheme.

The discussion of fermion representations in section \ref{sec:fermions}
suggests that a natural form of unification of gravity 
with all the other interactions would
consist of enlarging the local Lorentz group of the tetrad
formulations to $SO(3,11)$.
We will describe here a slightly more general case when
the enlarged group is some orthogonal group $SO(N)$
of suitable signature.

Since both gravity and Yang-Mills theories have strong geometrical
character, it is best to start the description of these unified theories
from the basic geometrical structures.
We again have an `internal'' vector bundle $E$ over spacetime,
but now its fibers have dimension $N>4$, while the base manifold $M$ remains  four dimensional.
As in the usual Einstein-Cartan formulation we assume that there is an ``internal'' metric $\eta$ in $E$, so that the group of ``vertical'' gauge transformations is $SO(N)$ of appropriate signature.

One clearly cannot assume anymore that $TM$ is isomorphic to $E$.
The strongest possible statement that one can make is that
$TM$ is a subbundle of $E$.
This amounts to the existence of a vectorbundle morphism $\theta$ 
(still called soldering form) of maximal rank (namely four).
Equation (\ref{tet}) still makes sense and says that the
metric on spacetime is the pullback of the internal metric
by the soldering form. 
Without loss of generality we can assume that the last four
elements of the basis $\{e_I\}$ in $E$ are in the image of $\theta$,
and the others are in the orthogonal complement.
In fact, we can choose them to be the images under $\theta$
of a tetrad in $TM$ for the pullback metric.
In such a basis the soldering form has components
\begin{equation}
\label{theta-gen}
\theta=\left[
\begin{array}{c} 0 \\
\mathbf{1}_4
\end{array}
\right].
\end{equation}

The connection field of this generalised Einstein-Cartan formalism is an ${\mathfrak so}(N)$ valued one-form that, in the same basis described above, can be represented in $(4,N-4)$ matrix
block notations
\begin{equation}\label{w-gen}
\omega_\mu=\left[
\begin{array}{cc}
\omega_\mu^{(N-4)}& K_\mu\\
-K_\mu & \omega_\mu^{(4)} 
\end{array}
\right].
\end{equation}
Here $\omega_\mu^{(4)}$ is an $SO(4)$ connection in the 4-dimensional subbundle $Im\theta\subset E$, and $\omega_\mu^{(N-4)}$ is an $SO(N-4)$ connection.

As in the Einstein-Cartan formalism,
we can define the connection in the tangent bundle
to be the pullback of $\omega_\rho$:
\be
\label{pullb}
\Gamma_\rho{}^\nu{}_\mu=
\theta_I{}^{\nu}\omega_\rho{}^I{}_J\theta^J{}_\mu
+\theta_I{}^\nu\partial_\rho\theta^I{}_\mu\ .
\ee
Note that $\theta^I{}_\mu$ does not have an inverse,
but we can define 
$\theta_I{}^\nu=\eta_{IJ}\theta^J{}_\sigma g^{\sigma\nu}$,
which has the property $\theta_I{}^\nu\theta^I_\mu=\delta^\nu_\mu$.
Equation (\ref{pullb})
can be obtained by multiplying (\ref{spinconn}) by
$\theta_I{}^\nu$.
It is therefore equivalent to a subset of those equations.

It is instructive to explore in some more detail the 
possible relations between the soldering form and the connection.
In Einstein-Cartan theory imposing the conditions of metricity and
vanishing torsion completely determines $\omega_{\rho IJ}$,
and then equation (\ref{spinconn}) fixes the 64 components of $\Gamma_\rho{}^\nu{}_\mu$ in terms of the 64 components
of $\omega_{\rho IJ}$.
Here the analogous relation is a bit more involved.

Imposing the antisymmetry (metricity) condition, $\omega_{\rho IJ}$
has $4\times N(N-1)/2$ independent components.
We can now impose equation (\ref{spinconn}),
which amounts to $16N$ conditions for $\omega_{\rho IJ}$
and for the 64 components of $\Gamma_\rho{}^\nu{}_\mu$.
This leaves us with $2N(N-1)+64-16N=2(N-4)(N-5)+24$ free functions.
We can further impose that $\Gamma_\rho{}^\nu{}_\mu$
be torsion-free, which amounts to 24 equations.
Altogether we remain with $2(N-4)(N-5)$ free functions,
which is just the number of components of
the internal $SO(N-4)$ YM field $\omega_\mu^{(N-4)}$.
In fact, the conditions of metricity and vanishing torsion
have entirely fixed the connection in $TM$, which is given
just by the Christoffel symbols of the pullback metric,
the spacetime components $\omega_\mu^{(4)}$ 
are the corresponding spin connection and
the mixed components $K_\mu$ have been forced to vanish.
Thus a theory with a dynamical $SO(N)$ connection
and soldering form, on which we impose by hand the constraints
(\ref{spinconn}) and absence of torsion, has the same
degrees of freedom as gravity coupled to an $SO(N-4)$ YM field.

Note that the case $N=5$ is a bit special, as was pointed out in \cite{Chamseddine:2010rv}, \cite{Chamseddine:2013hwa}: the preceding counting gives zero free functions.
This is simply because the normal to the tangent bundle
is one-dimensional, and the $SO(1)$-connection 
$\omega_\mu^{(1)}$ has a single component $\omega_{\mu 55}$
which must be zero by antisymmetry.
Furthermore, in this case the number of the components of the generalised frame is $4\times 5=20$. But the dimension of the gauge group has also increased as compared to $N=4$ case, and it is now $10$. The number of non-gauge components in the frame, which is the total number of components minus the dimension of the gauge group, is therefore still $10$, and so such a theory is effectively
just a theory of gravity.

In the following we will not impose (\ref{spinconn}) as
a constraint. Instead, we will see that it arises dynamically as a natural
property of the theory at low energy.
It is clear from the preceding discussion that the order parameter for this
gravity-Yang-Mills unification is the soldering form $\theta$ \cite{percacci1,percacci2}.
If it vanishes, then all the internal directions are equivalent.
When it has maximal rank, four of the internal directions have
a special character: they can be identified with the tangent
spaces to spacetime, while the others remain genuinely internal.
It is therefore the VEV of the soldering form that
separates the gravitational from the other interactions.

In this way we have fulfilled the first three points of the to-do list in section II.A. The hardest part is now to complete the fourth point, namely to write an $SO(N)$-invariant dynamics.

\subsubsection{Bosonic dynamics}
\label{sec:ecu_bose}

The first problem one encounters is that in the generalised context we are considering one can no longer write a Lagrangian in terms of differential forms. Indeed, we can wedge two copies of $\theta^I$ with the curvature $F^{IJ}$ to produce a four-form. But this leaves us with the problem of contracting the four internal indices in some way that does not produce a trivial theory. There seems to be no $SO(N)$-invariant way of doing this, unless one introduces other fields, as in the next subsection that discusses the coupling to fermions. This may be possible, but has not been explored. The only explored option is to abandon the idea of working with differential forms. This is a big departure from the Einstein-Cartan philosophy, but it appears that there is no other way forward if one is to pursue this unification scenario without introducing more fields. 

So, to write a Lagrangian we will take into account that when $N=4$ there are two equivalent ways of writing the Einstein-Cartan action. One is by using differential forms, as in (\ref{action-EC}). 
The other is by using the inverse vierbein as in (\ref{tetac}).
In the present context, the soldering form $\theta^I{}_\mu$
does not have an inverse, but one can use the internal metric 
$\eta_{IJ}$ and the induced spacetime metric $g_{\mu\nu}$,
assumed nondegenerate, to contract the indices. The action is 
\be\label{action-CM}
S[\theta^I{}_\mu, \omega_\mu^{IJ}] = \frac{1}{16\pi G} \int \sqrt{g} \left(\theta^I{}_\mu\theta^J{}_\nu 
g^{\mu\rho}g^{\nu\sigma}R(\omega)_{\rho\sigma IJ} - 2\Lambda\right). 
\ee
There are now two ways forward with this Lagrangian. One way, explored recently in \cite{Chamseddine:2010rv}, \cite{Chamseddine:2016pkx} is to impose the frame - connection compatibility equation (\ref{spinconn}) non-dynamically. As we discussed above, this equation fixes all the components of the connection in terms of the derivatives of the generalised frame, apart from the $\omega_\mu^{(N-4)}$ components. These components are then to be interpreted as the YM gauge fields. Then for $N=4,5$ this theory, after the connection is determined from (\ref{spinconn}) and is substituted back into the action, gives the Einstein-Hilbert Lagrangian that depends only on the metric, see \cite{Chamseddine:2010rv} for a further discussion. For $N>5$ one adds to the action terms quadratic in the curvature to generate the $F^2$ kinetic term for the YM fields, see \cite{Chamseddine:2016pkx}. 

The other possible way forward is to try to keep the full connection $\omega$ as an independent field, and let its relation to the metric arise dynamically, as is the case in the usual first-order formalism.
One possible way to do this is to drop the requirement that the action must be first-order in derivatives and add other types of terms. When one thinks of the most general possible action
for $\omega_\mu{}^{IJ}$ and $\theta^I_\mu$,
the most natural terms are quadratic in curvature
and in the total covariant derivative 
of the soldering form.
The effect of the latter terms is to conspire with the
action (\ref{action-CM}) to generate masses for the
$\omega_\mu^{(4)}$ and $K_\mu$ components of the connection
via the Higgs mechanism.\footnote{More precisely, what becomes massive is not 
$\omega_\mu^{(4)}$
but rather the difference between $\omega_\mu^{(4)}$
and the Levi-Civita connection constructed
with the soldering form \cite{percacci2}.}
The same effect is obtained if instead of the
total covariant derivative one employs
the covariant exterior derivative of the soldering form,
which is its antisymmetric part.
(It is given by equation (\ref{torsion}), where now $I,J=1,\ldots,N$.)
The terms quadratic in curvature  
produce the YM $F^2$-type kinetic terms for the part $\omega_\mu^{(N-4)}$ of the connection that is left massless. 

This construction is a close analog of unification
in the sense of particle physics, with non-linearly realized
order parameter, as discussed in section II.A.
There, the description in terms of nonlinearly realized fields
is the low-energy approximate theory describing the
physics below the scale of the Higgs VEV.
In the case of gravity this would presumably be the Planck scale.
Depending on dimensionless parameters appearing in the Lagrangian,
the mass of the connection may be comparable to or smaller than 
the Planck scale.
If one looks at this theory at scales much below the mass
of the gauge fields, the latter will appear to be dynamically frozen at their VEV.
This is the same as imposing the condition  (\ref{spinconn}) as a constraint.
Thus, the scheme recently discussed in \cite{Chamseddine:2016pkx} can be viewed as a low energy approximation of the theory discussed in \cite{percacci1,percacci2}, which in turn should be viewed as a low energy approximation of some more fundamental theory. 

At the classical level it is also consistent to
think of the theory with the constraint (\ref{spinconn})
as a gauge-invariant description of a massive connection,
independent of considerations of energy scales.
It is amusing to note that essentially the same logic has
also been used recently in the case of GUTs to justify the
absence of proton decay \cite{Karananas:2017mxm}.

We stress once more that the action involves 
a non-degenerate metric
and therefore only makes sense in the low-energy (broken) phase
of the theory. From the point of view of the criteria for unification
spelled out in section II.A, we fall short of having
a completely satisfactory dynamics.

To summarise, what appears to be the most serious drawbacks of this unification scenario are the departure from the first order formalism, and a related departure from the requirement that the Lagrangian be written in terms of differential forms.  A related drawback is that this unified theory can only describe the dynamics of the broken phase. 

We note that some attempts to provide a dynamical justification for
the non-vanishing VEV of the soldering form, by means of
a self-consistent, bi-metric dynamics, were made in
\cite{percacci1,percacci2,flops,flop2,Floreanini:1993na}.
While bi-metric dynamics can be seen as aesthetically 
unpleasant, they have been used extensively recently in discussions of massive gravity \cite{Hinterbichler:2011tt,deRham:2014zqa}
and also appear in approaches to asymptotic safety
\cite{reuterrev,perbook2}.

\subsubsection{Fermion dynamics}

When one has a bosonic dynamics that can explain the generation
of a nonzero VEV for the soldering form, then the formulation
of a suitable fermionic dynamics satisfying all the criteria
of section II.A is relatively straightforward.
Since the spinorial representations depend on the dimension,
we consider here the special case of the $SO(3,11)$ unification
mentioned above \cite{gravigut}.
We start from the Clifford algebra of $SO(3,11)$, generated by gamma matrices
$\gamma^I$ (latin indices $I,J$ now run from $1$ to $14$), satifying 
$\{\gamma^I,\gamma^J\}=2\eta^{IJ}$.
The $SO(3,11)$ covariant derivative acting on Majorana-Weyl spinors is
\begin{equation}
D_\mu\psi_{L+}
=\left(\partial_\mu+\frac{1}{2}\omega_\mu^{IJ}\Sigma_{L\,IJ}^{(3,11)}\right)\psi_{L+}
\end{equation}
where $\Sigma_{IJ}^{(3,11)}={\tiny \frac{1}{4}}[\gamma_I,\gamma_J]$
are the generators of $SO(3,11)$ and $\Sigma_{L\,IJ}^{(3,11)}$ 
their restriction to the (left-handed) Majorana-Weyl representation.
We also define the covariant differential $D$, mapping spinors to spinor-valued
one forms: $D\psi_{L+}=(D_\mu\psi_{L+})dx^\mu$.
There is an intertwiner $A$ mapping the spinor representation to its 
hermitian conjugate: $\Sigma_{IJ}^\dagger A=-A \Sigma_{IJ}$.
Therefore the quadratic form
\begin{equation}
\label{eq:quad}
\psi_{L+}^\dagger (A\gamma^I)_LD\psi_{L+}
\end{equation}
is manifestly a vector under $SO(3,11)$ and a one form under diffeomorphisms. 
Then, to construct an $SO(3,11)$-invariant action, we introduce an auxiliary field
$\phi_{IJKL}$ transforming as a totally antisymmetric tensor. The
action is
\begin{equation}
\label{eq:action}
\mathcal{S}=\int\psi_{L+}^\dagger (A\gamma^I)_LD\psi_{L+}\,\wedge\theta^J\wedge\theta^K\wedge\theta^L \,\phi_{IJKL} \,.
\end{equation}
 
The breaking of the $SO(3,11)$ group to $SO(10)$
is induced by the VEV of two fields: the soldering one-form
$\theta^I{}_\mu$ and the four-index antisymmetric field $\phi_{IJKL}$. We
assume that the VEV of $\phi_{IJKL}$ is $\epsilon_{mnrs}$, the
standard four index antisymmetric symbol, in the Lorentz subspace 
(spanned by indices $m,n=1,2,3,4$), and
zero otherwise.\footnote{We note that the field $\phi_{IJKL}$
also appears in BF reformulations of General Relativity, as reviewed above.}
The VEV of the soldering form on the other
hand has maximal rank (four) and is also nonvanishing only in the
Lorentz subspace:
\begin{equation}
\left\{\begin{array}{l}
\phi_{mnrs}=\epsilon_{mnrs}\\
\phi_{IJKL}=0 \quad\mathrm{otherwise}
\end{array}
\right.
\quad
\left\{\begin{array}{l}
\theta^m{}_\mu=M e^m{}_\mu\\
\theta^I{}_\mu=0 \quad\mathrm{otherwise}
\end{array}
\right.
\label{eq:VEVs}
\end{equation}
where $e^m{}_\mu$ is a vierbein, corresponding to some solution of the
gravitational field equations which we need not specify in this
discussion (below we will choose $e^m{}_\mu=\delta^m_\mu$) and $M$ can
be identified with the Planck mass.

Then, the action for fluctuations around this VEV reduces to the standard 
action for a single $SO(10)$ family in flat space:
\begin{equation}
\label{eq:action}
\int d^4x\,\eta^\dagger\sigma^\mu \nabla_\mu\eta \,,
\end{equation}
where now
$\nabla_\mu=D^{(10)}_\mu=\partial_\mu+\frac{1}{2}A_{\mu\,(10)}^{ab}\Sigma^{(10)}_{ab}+\frac{1}{2}A_{\mu\,(3,1)}^{mn}\Sigma^{(3,1)}_{mn}$
is the Lorentz- and $SO(10)$-covariant derivative. Note that this action contains
the standard kinetic term of the fermions, and the interaction with
the $SO(10)$ gauge fields, which at this stage can still be assumed to be massless. 

We note that a scalar field $\phi_{IJKL}$ is reminiscent of what is needed in the MacDowell-Mansouri scenario, to be discussed later. Indeed, a field of this type can also be used as the ``compensator'' field in the Lagrangian $\phi_{IJKL} F^{IJ} \wedge F^{KL}$. So, it may be that the fermionic Lagrangian described above should also be used in the context of the MacDowell-Mansouri type unification, see below. This has not been explored. 

In summary, we see that it is possible to write 
an $SO(3,11)$-invariant action for the fermions
that reduces to the correct Lorentz- and $SO(10)$-invariant 
action in the broken phase. The most difficult part is thus to get the satisfactory mechanism for the symmetry breaking in this context. 
Besides the explicit constructions discussed
in the preceding section, we mention the possibility
of a dynamical symmetry breaking in a purely spinorial theory.
This has been explored to some extent in 
\cite{Hebecker:2003iw,Wetterich:2003wr}.

\subsubsection{Graviweak unification}
\label{sec:graviweak}

\def\sltwoc{SL_\C(2)}
\def\sutwo{SU(2)}
\def\sutwol{\sutwo_L}
\def\sothreeone{SO(3,1)}
\def\sothreeonec{SO_\C(3,1)}
\def\sofourc{SO_\C(4)}
\newcommand\al{\alpha}
\newcommand\sh{\hat\sigma}
\newcommand\one{{\mathbf{1}}}

Less ambitious than the ``graviGUT'' discussed in
the preceding sections, this is a theory unifying
gravity with the weak interactions in a complex orthogonal group
\cite{Nesti:2007ka}.
It is easiest to motivate this sort of unification
if one starts from a simplified setting where
the right-handed fermions are absent and
the left-handed ones are doublets of $\sutwol$.
We ignore strong interactions. 
Since the fermions are
complex, they automatically carry a representation of the
\emph{complexified} Lorentz and weak groups.  The algebra of the
complexified Lorentz group $\sothreeonec\equiv\sofourc$ consists of real linear
combinations of the rotation generators $L_j$, the boost generators
$K_j$ and their purely imaginary counterparts $iL_j$ and $iK_j$.  In
the case of the chiral fermion fields, 
the physical rotations and boosts are realized by the
generators $M^+_j=L_j+iK_j$ and $iM^+_j$ respectively, which together
generate a group $\sltwoc_+$. The generators $M^-_j=L_j-iK_j$ of
$\sofourc$ commute with the $M^+_j$ and can therefore be
identified with physical operations on spinors that have nothing to do
with Lorentz transformations.  
In this simplified chiral model we can
identify $\sltwoc_+$ with the Lorentz group, and the group generated
by the $M^-_j$ with the weak isospin gauge group $\sutwol$.  The
generators $iM^-_j$ are related to the weak isospin generators in the
same way as the boosts are related to the rotations, therefore we can call them ``isoboosts'' and we can call the group $\sltwoc_-$
generated by $M^-_j$ and $iM^-_j$ the ``isolorentz group''.  It is
just the complexification of the isospin group.  
The group $\sothreeonec\equiv\sofourc=\sltwoc_+\times \sltwoc_-$,
which contains both Lorentz and isolorentz transformations, 
is called the ``graviweak'' group.
Since this group is a direct product, it may seem that
no true unification has been achieved in this way.  However, it is
both mathematically and physically different 
to have a gauge theory of
the group $\sofourc$, with a single coupling constant, and of the
group $\sltwoc\times\sltwoc$, which in general has two.

We shall use the following conventions regarding the indices:
$a,b=1,2,3,4$
are indices in the vector representations of the real
$SO_\R(3,1)\subset SO_\C(3,1)$ generated by $(L_j,K_j)$, while 
$m,n=1,2,3,4$ 
are indices in the vector representations of
$\sltwoc_+$ generated by $(M_j^+,iM_j^+)$, and
$u,w=1,2,3,4$ 
are indices in the vector representations of
$\sltwoc_-$ generated by $(M_j^-,iM_j^-)$.

In this theory one can write the action in terms of
differential forms.
The order parameter is a generalized soldering form
$\theta_\mu^{\bar a a}$, which can also be written as $\theta_\mu^{mw}$.
Denoting $A_\mu{}^a{}_b$ and $\bar A_\mu{}^{\bar a}{}_{\bar b}$ the graviweak gauge field and its conjugate, 
the generalized torsion is
\be
\Theta_{\mu\nu}^{\bar aa}=
\partial_\mu\theta^{\bar aa}{}_\nu-
\partial_\nu\theta^{\bar aa}{}_\mu+
\bar A_\mu{}^{\bar a}{}_{\bar b}\theta^{\bar b a}{}_\nu+
A_\mu{}^a{}_b\theta^{\bar a b}{}_\nu -
\bar A_\nu{}^{\bar a}{}_{\bar b}\theta^{\bar b a}{}_\mu-
A_\nu{}^a{}_b\theta^{\bar a b}{}_\mu 
\ee
and the curvature two-form is
\bea
R_{\mu\nu}{}^{\bar aa\,\bar bb}&=&
R_{\mu\nu}{}^{ab}\delta^{\bar a\bar b}
+\bar R_{\mu\nu}{}^{\bar a\bar b}\delta^{ab}\\
\label{eq:R}
R_{\mu\nu}{}^a{}_b&=&
\partial_\mu A_\nu{}^a{}_b
-\partial_\nu A_\mu{}^a{}_b+
A_\mu{}^a{}_cA_\nu{}^c{}_b 
- A_\nu{}^a{}_cA_\mu{}^c{}_b\,.
\eea
 
With these fields we can define a generalized Einstein-Cartan
action, which contains terms
\bea
\label{eq:EHsymm}
S_{EC}&=&\frac{g_1}{16\pi}\int 
 R^{\bar aa \,\bar bb}\wedge \theta^{\bar cc}\wedge \theta^{\bar dd}\,\epsilon_{(\bar aa)(\bar bb)(\bar cc)(\bar dd)}\\
\label{eq:torsionsymm}
S_\Theta&=&a_1\int\!\bigg[ 
  t_{\bar ee}^{\bar aa\,\bar bb}\,\Theta^{\bar ee} 
  +(t^2)\,\theta^{\bar aa}\wedge \theta^{\bar bb}
           \bigg]\wedge\theta^{\bar cc}\wedge \theta^{\bar dd}
 \epsilon_{(\bar aa)(\bar bb)(\bar cc)(\bar dd)}
\eea
where $t_{\bar ee}^{\bar aa\,\bar bb} $ are zero-form auxiliary fields.
Eliminating them, the second term is quadratic in torsion.
Similarly
\bea
\label{eq:R2symm}
S_2&=&\frac1{g_2^2}\int 
\!\bigg[r_{\bar ee\,\bar ff}^{\bar aa\,\bar bb}\,R^{\bar ee \,\bar ff}  
 + (r^2)\,\theta^{\bar aa} \wedge\theta^{\bar bb}\bigg]
\wedge\theta^{\bar cc}\wedge 
\theta^{\bar dd}\epsilon_{(\bar aa)(\bar bb)(\bar cc)(\bar dd)}\,.
\eea
is quadratic in graviweak curvature after eliminating the
auxiliary fields $r_{\bar ee\,\bar ff}^{\bar aa\,\bar bb}$.

The equations of motion of this action admit Minkowski space
as a solution.
We shall refer to this solution as the VEV.
It is given by $\langle A_\mu{}^a{}_b\rangle=0$ and 
$\langle\theta^{m4}_\mu\rangle=M\delta^m_\mu$ and 
$\langle\theta^{mu}_\mu\rangle=0$ for $u=1,2,3$,
where $M$ is a mass parameter. 
This VEV breaks the original group in the correct way to provide
global Lorentz and local weak (isospin) gauge invariance: the $(+)$
part of the $\sofourc$, corresponding to the Lorentz
generators, and the imaginary part of the $(-)$ generators (the isoboosts) do not leave the VEV invariant, and therefore are broken.  
Thus, the only unbroken subgroup of the
original gauge group is the weak $\sutwol$.  
In addition, the VEV $\theta^m_\mu=\delta^m_\mu$ is invariant under the global diagonal $\sothreeone$.  This is the
usual Lorentz group.
Notice that the VEV has selected $\sltwoc_+$ for
soldering with the spacetime transformations, and accordingly the
signature of the resulting metric is Minkowskian.

In order to describe in a covariant fashion also 
non-flat geometries with weak curvature we can
consider backgrounds of the form:
\be
\label{eq:VEV}
\langle\theta^{m4}_\mu\rangle=M e_\mu^m(x)\ ,
\langle\theta^{mu}_\mu\rangle=0\ \mathrm{for}\ u=1,2,3
\ee
where $e_\mu{}^m$ are now ordinary, real vierbeins connecting the internal Lorentz vector index $\mu$ 
to the internal vector index $m$.  
Moreover, using the $\sofourc$ invariant product $\delta_{ab}$,
one can define a metric 
$g_{\mu\nu} =\theta_\mu^{\bar aa}\theta_\nu^{\bar bb}
\delta_{ab}\delta_{\bar a\bar b}= e_\mu^m e_\nu^n\,\eta_{mn}$, 
where in the last step we used (\ref{eq:VEV}).

If the metric is slowly varying we can neglect the action $S_2$.
In deriving the equations of motion (EOMs) 
for the other part of the action it is convenient to split
the connection and curvature in selfdual and antiselfdual parts,
converting the graviweak indices $(\bar aa)$ to Lorentz indices
$m,n\ldots$ and isolorentz indices $u,v,\ldots$.  
Then, the EOMs for the isolorentz (anti-selfdual)
connection are identically satisfied by the
VEV~(\ref{eq:VEV}), while the equation for the Lorentz (selfdual)
connection imply that the standard gravitational torsion vanishes:
\be
\Theta_{\mu\nu}^{m}\equiv
\partial_\mu e^{m}{}_\nu-\partial_\nu e^{m}{}_\mu+
\omega_\mu{}^{m}{}_{n}e^{n}{}_\nu
+\omega_\mu{}^{m}{}_{n}e^{n}{}_\mu=0\,.
\ee
This fixes $\omega_\mu{}^m{}_n$ to be the Levi-Civita connection of
$e_\mu^m$.  On the other hand the equation relative to $\theta_\mu^{mu}$
produces the Einstein equations for the background~$e_\mu^m$.  Thus, if
$e_\mu^m$ is a solution of Einstein's equations in vacuum,
then~(\ref{eq:VEV}) 
yields a solution of the equations of motion of this theory.

One can understand better the dynamics of the gauge fields by
inserting the VEV~(\ref{eq:VEV}) in the action and neglecting
interaction terms.  The generalized actions~(\ref{eq:EHsymm})
and~(\ref{eq:torsionsymm}) become
\be
S_{EC}+S_\Theta \to  \int {\rm d}^4x\,\sqrt{g} \Big[
\frac{g_1}{16\pi} M^2 R+ 4a_1M^2
\left(\Theta_{\mu\nu}^m \Theta^{\mu\nu}_m 
+ 10\, K^{j}_\mu\,K_{j}^\mu \right) 
\Big]\,.
\ee
Thus one should identify the Planck mass as $M_{PL}^2=g_1M^2$.  Then,
this shows that the isoboost gauge fields $K_\mu^j$ acquire mass at the
Planck scale. As discussed in the introduction, also the
spin-connection $\omega_\mu^j$, which is contained in 
$\Theta_{\mu\nu}^m$ and $R$, becomes massive.  
This can be seen most clearly for the constant
background $e_\mu^m=\delta_\mu^m$; in curved backgrounds, it will
generate masses for the fluctuations of $\omega$ around the Levi-Civita
connection of $e_\mu^m$.  The $W$ boson remains massless.

The action $S_2$ modifies the equations for the VEV, but flat space is still a solution.
Using (\ref{eq:VEV}) and eliminating the auxiliary fields, 
the action $S_2$
reduces to a term quadratic in the gravitational curvature plus the standard Yang-Mills actions for the weak gauge fields:
\be
S_2 \to \frac1{g_2^2}\int \!{\rm d}^4x \sqrt{g} \,\bigg(- R_{\mu\nu}^j R^{\mu\nu}_j
- W_{\mu\nu}^j W^{\mu\nu}_j-K_{\mu\nu}^{j} K^{\mu\nu}_{j}\bigg )\,.
\ee
Above the breaking scale, the gravi-weak symmetry manifests itself in the equality of the coefficients of all the three terms, while below the Planck scale the isoboosts and the spin connection are massive and decoupled. 

One should point out that the
equations admit also the solution $\langle\theta\rangle=0$.  This
corresponds to an ``unbroken'' phase in which there is no distinction
between gravitational and weak interactions.  Since the metric is
quadratic in $\theta$, one expects this symmetric phase to be also
``topological''.  The dynamical mechanism which favours the phase with
nondegenerate metric is outside this picture,
but both phases at least appear as admissible solutions.

One can modify the theory to includes also the
right-handed fermions and the strong interactions \cite{Nesti:2007ka}, see also \cite{Alexander:2007mt}. This theory has been used for cosmological applications
in \cite{Das:2013xha,Das:2014xga,Laperashvili:2014xea,Das:2015usa}.

\subsection{MacDowell-Mansouri type unification}

The MacDowell-Mansouri action for General Relativity (\ref{action-MM}) is based on the DeSitter or Anti-DeSitter gauge group, and possibly an explicit vector field that breaks the symmetry to Lorentz as in (\ref{action-SW}). It has been realised early on that other gauge groups can be considered. Indeed, one of the motivations of the original MacDowell-Mansouri paper \cite{MacDowell:1977jt} was a simple construction of supergravity along these lines, with a supergroup replacing the DeSitter or Anti-DeSitter gauge groups. However, it took many years before any serious investigation as to other possibilities was carried out. 

In three spacetime dimensions the MacDowell-Mansouri (or Cartan) trick of putting together the frame and the spin connection leads to the Chern-Simons description \cite{Witten:1988hc}. It is interesting to remark that in this Chern-Simons context the procedure of ``enlarging the gauge group'' from the $SU(2)$ that is needed for gravity to higher rank groups has been studied extensively. It turns out that the theories one gets this way are related to higher spin theories, see e.g. \cite{Campoleoni:2012hp}.

In the setting of four dimensions, the paper \cite{Zlosnik:2016fit} studies MacDowell-Mansouri-type theory with the conformal group $SU(2,2)\sim SO(2,4)$. We review their construction below. 

Other papers on extended MacDowell-Mansouri formalism include: An interesting paper \cite{Westman:2013mf} studying Stelle-West-type actions with a potential term for the compensator field instead of the Lagrange multiplier term. It is shown that the result is a variant of scalar-tensor theory of gravity. 
Lisi considered a MacDowell-Mansouri-type action for the gauge group as large as $E_8$, attempting also to include fermions as components of some superconnection. 
We will discuss this in section V.G.
Additional work on ``enlarging the gauge group'' in the context of MacDowell-Mansouri formulation is \cite{Durka:2011nf}.

\subsubsection{MacDowell-Mansouri-type theory with conformal group}

As in \cite{Zlosnik:2016fit}, let $A,B,\ldots$ be 6-dimensional indices so that an object $V^A$ is in the six-dimensional defining representation of $SO(2,4)$. The connection is then a Lie algebra-valued one-form $A^{AB}=A^{[AB]}$. Its curvature is $F^{AB}$. To construct the Lagrangian with need an object $w^{AB}$ with two indices to contract with $\epsilon_{ABCDEF} F^{CD}\wedge F^{EF}$. This is in contrast with a one-index object in the Stelle-West version (\ref{action-SW}) of the usual MacDowell-Mansouri formalism. The authors of \cite{Zlosnik:2016fit} start with $W^{AB}$ in a general orbit under $SO(2,4)$, but then quickly specialise to vectors of the form
\be
W^{AB} = \left( \begin{array}{cc} 0 & 0 \\ 0 & \bar{\phi}\, \epsilon^{ab} \end{array} \right),
\ee
where the upper-diagonal block is $4\times 4$, and the indices $a,b$ take two values. It is assumed that $\bar{\phi}$ is a constant. Similar decomposition of the connection is
\be\label{ZW-conn}
A^{AB} = \left( \begin{array}{cc} w^{IJ} & E^{Ia} \\ -E^{Ia} & c \epsilon^{ab} \end{array} \right).
\ee
Here $I,J,\ldots$ are 4-dimensional internal indices. It is immediately clear that the novelty in the ``enlarged gauge group'' case is that there is now not one but two fields $E^{I \,1,2}$ that can play the role of the frame field. Introducing the sum and difference linear combinations $e^I, f^I$ out of $E^{I \, 1,2}$ the authors obtain the ``broken phase'' action of the following form
\be\label{action-ZW}
S[w,c,e,f]= \int \alpha \epsilon_{IJKL} e^I f^J R^{KL} + \beta e_I f_J R^{IJ} + \gamma \epsilon_{IJKL} e^I f^J e^K f^L + \mu e_I f^I e_J f^J + \xi e_I f^I dc,
\ee
where $\alpha,\beta,\gamma,\mu,\xi$ are all constants whose values are related to $\bar{\phi}$. Apart from the last term containing $dc$, this is the action of the type considered in \cite{Hinterbichler:2012cn} in the context of bi-metric gravity. It may therefore describe a massless and a massive graviton. Unlike \cite{Hinterbichler:2012cn}, however, there is an additional symmetry $e^I \to e^{\alpha(x)} e^I, f^I \to e^{-\alpha(x)} f^I$ in (\ref{action-ZW}). For an analysis of the perturbative spectrum of this theory see \cite{Zlosnik:2016fit}.

\subsubsection{General case and difficulties}

There is clearly a generalisation of the above construction to arbitrary $SO(N)$ gauge group. One wants to break this gauge group to $SO(4)\times SO(N-4)$. In the Stelle-West-type approach this breaking will be carried out by a compensator field, which is totally anti-symmetric in $N-4$ indices, an analog of $v^a$ in (\ref{action-SW}) or $W^{AB}$ in the previous section. It is also clear that there is an analog of the decomposition (\ref{ZW-conn}) in the general case, with the off-diagonal components of this matrix playing the role of a set of tetrad-like fields. The unbroken symmetry group $SO(N-4)$ acts by mixing these tetrads. In general this gives a version of multi-tetrad theory of \cite{Hinterbichler:2012cn}, but with an additional gauge symmetry. It would be interesting to study these theories better to understand their viability. 

One obvious difficulty of the models of this type is that, while YM-like fields valued in $SO(N-4)$ do get generated, the type of Lagrangians that one would naturally write in this formalism only gives first-order kinetic terms for these fields, not second order. So, one will never get the YM $F^2$ terms from first-order Lagrangians of the sort discussed. This is typical of all first-order formulations. The desired $F^2$ terms may in principle be obtained by integrating out some other fields, in this case the components of the generalised frame field. This, however, seems unlikely given that the frame fields are one-forms, and to get $F^2$ terms one expects to integrate out two-forms, as we will see in the context of BF-type formalisms below. This issue, however, needs to be studied better. 

Another property that we see without any analysis is that the set of tetrads that one will get from these models will be charged with respect to the unbroken gauge group $SO(N-4)$. While an $SO(N-4)$-invariant combination can be formed to play the role of the ``physical'' metric, the interpretation of the other $SO(N-4)$ charged components remains obscure. Thus, it is far from clear that the spectrum of propagating modes in these theories will resemble what one wants to get.

Another issue with this unification scenario is that nothing in principle prevents one from taking an arbitrary gauge group containing $SO(4)$. Indeed, there is nothing in this gauge-theoretic scheme that forces us to stick to orthogonal groups. However, fermions seem to suggest that the relevant unification group is an orthogonal group. 

Given these difficulties, the set of models that can be obtained this way does not seem to be too promising for the purpose of unification of gravity with other SM bosonic fields. 

\subsection{BF type unification}

Historically, gravity was first reformulated as constrained BF theory by Plebanski \cite{Plebanski:1977zz}. His paper contained both the chiral and non-chiral versions. It seems that it was Robinson \cite{Robinson:1994yy} who first thought of unification in the framework of this formalism, even though his paper is based on the chiral Plebanski formulation and so we postpone its treatment to the next section. Another early paper on the subject of unification is \cite{Peldan:1992iw}, but it is again about the chiral formalism, and moreover uses a non-manifestly covariant Hamiltonian framework, so we refrain from reviewing it in this work. 

The first paper studying the unification based on non-chiral BF formalism for gravity was \cite{Smolin:2007rx}, with the motivation for modifying the non-chiral Plebanski action coming from the work on ``deformations of GR'' \cite{Krasnov:2006du} by one of the present authors. This unification scenario was further developed in \cite{Smolin:2009bj}, \cite{Speziale:2010cf} and \cite{Lisi:2010td}, as well as in \cite{Beke:2011dp}, \cite{Beke:2011mu}. Another relevant paper that uses this formalism is \cite{Alexander:2012ge}. It is these non-chiral BF unification scenario developments that we will aim to review in this section. 

\subsubsection{Modified Plebanski}

The main idea of \cite{Krasnov:2006du} was to modify the chiral Plebanski theory by removing the constraints that the variation with the Lagrange multiplier field imposes. The idea of Smolin \cite{Smolin:2007rx} was to combine this with the ``enlarging the gauge group'' idea. 

In retrospect, one proceeds in two steps. First, the non-chiral Plebanski action (\ref{action-FdP}) is modified to
\be\label{action-FdP-mod}
S[B,w,\Psi] = \frac{1}{16\pi G} \int B_{IJ}\wedge F^{IJ}(w) - \frac{1}{2} \left( \Psi^{IJKL} +\left(\frac{\Lambda}{6}  + \alpha (\Psi^{IJKL})^2\right) \epsilon^{IJKL}\right) B_{IJ}\wedge B_{KL}.
\ee
The ``Lagrange multiplier'' field $\Psi^{IJKL}$ is still taken to be tracefree, but the variation with respect to this field no longer imposes a constraint on the 2-form field $B^{IJ}$. Rather, one gets a set of equations from which the field $\Psi^{IJKL}$ can be determined in terms of the components of $B^{IJ}\wedge B^{KL}$ matrix. 

As the paper \cite{Alexandrov:2008fs} showed, the modification (\ref{action-FdP-mod}) is not innocuous, as new propagating degrees of freedom are added in the process. The paper \cite{Speziale:2010cf} interpreted the arising theory as a bi-metric theory of gravity with $2+6$ propagating degrees of freedom corresponding to a massless and a massive graviton. Some further aspects of this theory were later studied in  \cite{Beke:2011dp}, \cite{Beke:2011mu}. 

\subsubsection{Unification by enlarging the gauge group}

The second step, which is the main idea of \cite{Smolin:2007rx}, was to enlarge the gauge group in (\ref{action-FdP-mod}) from $SO(4)$ to an arbitrary group containing $SO(4)$ as a subgroup. Let us for definiteness assume this larger gauge group to be an orthogonal group $SO(N)$, even though there is nothing in this unification scenario that restricts us to orthogonal groups. Let us keep using the letters $I,J,\ldots$ to denote the $N$-dimensional internal indices. One of the main points of \cite{Smolin:2007rx} is that there is a solution of the field equations of (\ref{action-FdP-mod}) that breaks the symmetry $SO(N)$ down to $SO(4)$ times the subgroup that commutes with it. This is similar to how a generalised tetrad in the Einstein-Cartan-type scenarios breaks the $SO(N)$ symmetry as in (\ref{theta-gen}). The breaking pattern will in general depend on the embedding of the unbroken $SO(4)$ into the full gauge group $SO(N)$ selected by the solution in question, as was emphasised and explored in \cite{Krasnov:2011hi} in the context of the unification based on chiral formalism. For simplicity, we assume that the symmetry breaking pattern is $SO(N)$ down to $SO(4)\times SO(N-4)$. If we use indices $a,b,\dots$ to denote the first four of the indices $I,J,\dots$, i.e. say $I=(a,i), a=1,\ldots,4, i = 5,\ldots N$, then the relevant solution can be described as
\be\label{B-vac}
B^{ab} = \frac{1}{2} \epsilon^{abcd} \theta^c \wedge \theta^d,
\ee
where $\theta^a$ is a four-dimensional frame field, and all other components of $B^{IJ}$ are zero. Thus, the background field configuration for the $B^{IJ}$-field selects a particular $SO(4)$ subgroup in the full gauge group. The background value of the field $\Psi^{IJKL}$ is taken to be zero, and the only nontrivial components of the background connection are $\omega^{ab}$, assumed to be the spin connection compatible with the frame field $\theta^a$, which 
in turn is assumed to be maximally symmetric, 
i.e. correspond either to DeSitter or Anti-DeSitter space, depending on the sign of $\Lambda$ in the action (\ref{action-FdP-mod}). 

The idea is then that the perturbations around the selected symmetry breaking background will describe gravitons as well as Yang-Mills fields for the unbroken gauge group $SO(N-4)$. Nobody seems to have analysed these perturbations carefully, such an analysis was only done in the chiral version of this unification scheme \cite{Krasnov:2011hi}. But the results of the analysis in the chiral case suggest that the following behaviour can be expected. The $SO(4)$ sector of the theory will describe some version of bi-metric gravity with $2+6$ propagating degrees of freedom. The  $SO(N-4)$ sector will describe Yang-Mills theory. The $F^2$ form of the action for these Yang-Mills field follows by integrating out the $B^{ij}$ components of the 2-form field, as well as the $\Psi^{ijkl}$ components of the $\Psi^{IJKL}$ field. Further, there are what can be referred to as off-diagonal components of all the fields charged with respect both the Lorentz group $SO(4)$ as well as the Yang-Mills gauge group $SO(N-4)$. These describe exotic fields, referred to as Higgs fields in \cite{Lisi:2010td}. 

We should point out that a slightly different action from (\ref{action-FdP-mod}) was considered in \cite{Lisi:2010td}, with up to cubic dependence on the analog of the field $\Psi^{IJKL}$, which in this paper is also taken to have some spacetime indices. But the overall logic remains unchanged. We should also point out that the paper \cite{Alexander:2012ge} suggested that in a certain parity asymmetric phase of the $SO(4)$ theory one of the two $SU(2)$'s of the $SO(4)$ can be interpreted as the gauge group that corresponds to gravity, while the other one gives the gauge group of weak interactions. 
This is similar to the idea of graviweak unification
discussed in section \ref{sec:graviweak}.

One of the main achievement of the discussed formalism is that it is first-order in derivatives, works in terms of differential forms, and successfully solves the problem of generating the $F^2$ terms for the Yang-Mills gauge fields. Indeed, these are obtained by integrating the relevant components of the two-form field from the original first-order action. However, the difficulty with this formalism is that, at least around the $SO(4)$ symmetric vacuum (\ref{B-vac}) the massless spin two particle arises together with its massive cousin. This seems undesirable. Given that all 6 polarisations of the massive graviton propagate, there is likely the ghost mode, even though this issue strongly depends on the reality conditions chosen for all the fields, and these are subject to debate. It is probably the appearance of this massive graviton mode that led to diminishing interest in this unification model. A possible way out was advocated in \cite{Alexander:2012ge}, and is to expand around a different, parity asymmetric background, but then one faces the problems of reality conditions, see next section. So, the status of this unification scheme at present is unclear. 

The other difficulty of this scenario is the appearance of fields that transform with respect to both the YM gauge group, as well as Lorentz group. Such fields are clearly undesirable, but it is possible that they arise as massive fields, and that this mass can be tuned to be large. This needs to be studied in more details. The final difficulty is that in any formalism that is based on 2-forms, not frame fields, a coupling to fermions is problematic. A possible such coupling, but in the context of the chiral theory, was described in \cite{Capovilla:1991qb}, but it is far from clear that this coupling survives the generalisation from (\ref{action-FdP}) to (\ref{action-FdP-mod}), as we will discuss in more details below. 

\subsection{Chiral unification}

The unification scenario that starts with the chiral Plebanski formalism (\ref{action-Pleb}) has been studied by one of the present authors and collaborators. The paper \cite{TorresGomez:2009gs} studied an $SU(3)$ model, linearising the action of the full theory around a solution that breaks the symmetry to $SU(2)\times U(1)$, and interpreting the arising excitations as gravitons, Maxwell field, as well as exotic ``Higgs'' fields. The paper \cite{TorresGomez:2010cd} considered a similar theory but with the gauge group $GL_\C(2)$ with what arises being a version of unified theory of non-linear electrodynamics and gravity. Both papers work with BF-type formalism. The second of these papers also analyses the non-linear aspects of unification and in particular solves the spherically symmetric problem in the full non-linear theory. 

The paper \cite{Krasnov:2011hi} is about the same unification scheme, but the starting point is a pure connection action with an arbitrary gauge group. It is shown that there are in general many different possible vacua for the theory, each vacuum being determined by the embedding of the gravity gauge group $SU(2)$ into the full gauge group. The spectrum of excitations one finds around the vacuum strongly depends on this embedding. 

We start our review of the chiral models with the simplest and possibly the most attractive model of this type, the one described in \cite{Robinson:1994yy}.

\subsubsection{$GL_\C(2)$ Plebanski-type Einstein-Maxwell unified theory}
\label{sec:Robinson}

The reference \cite{Robinson:1994yy} considers a theory of exactly the same type as (\ref{action-Pleb}) but with $GL_\C(2)$ gauge group instead of $SL_\C(2)$. We will write this action in $SO(3)$ indices, similar to (\ref{action-Pleb}), and add another direction to the Lie algebra to represent the $U(1)$ gauge group. Thus, let the index $I=(i,4), i=1,2,3$ take four values. The action is
\be\label{action-Robinson}
S[A,B,M] = \int B^I F^I - \frac{1}{2} M^{IJ} B^I B^J + \mu_1({\rm Tr}_{SO(3)}(M) - \Lambda) + \mu_2( {\rm Tr}_{U(1)}(M) - k).
\ee
Thus, we enlarged the gauge group of the Plebanski formulation by adding $U(1)$, and further added another trace condition on the matrix that appears in front of the 4-form $B^I\wedge B^J$. Here the traces are ${\rm Tr}_{SO(3)}(M)\equiv M^{ij} \delta_{ij}$ and ${\rm Tr}_{U(1)}(M)\equiv M^{44}$. Thus, the constraints present in (\ref{action-Robinson}) require that the matrix $M^{IJ}$ is of the form
\be\label{Rob-M}
M^{IJ} = \left( \begin{array}{cc} \Psi^{ij} + \frac{\Lambda}{3} \delta^{ij} & \phi^i \\ \phi^i & k\end{array} \right).
\ee

To see that the theory (\ref{action-Robinson}) is equivalent to Einstein gravity coupled to Maxwell let us write everything in $SO(3)$ plus $U(1)$ components. We have
\be\label{action-Rob-comp}
S[A,B,\Psi,\phi] = \int B^i F^i  + B^4 F^4 - \frac{1}{2} \left( \Psi^{ij} + \frac{\Lambda}{3} \delta^{ij}\right)B^i B^j   - \frac{1}{2} k B^4\wedge B^4 - \phi^i B^i B^4,
\ee
where we have used (\ref{Rob-M}). 

Now, the $SO(3)$ sector is unchanged as compared to (\ref{action-Pleb}) and continues to describe General Relativity. Varying with respect to $\phi^i$ gives
\be
B^i \wedge B^4=0,
\ee
which implies that $B^4$ is a purely anti-self-dual 2-form. Using this fact, $B^4$ can be integrated out from the action using its field equation
\be\label{B4-F4}
k B^4 = (F^4)_{asd}.
\ee
This gives the following action
\be
S[A^i,B^i,\Psi, A^4] = \int B^i F^i - \frac{1}{2} \left( \Psi^{ij} + \frac{\Lambda}{3} \delta^{ij}\right)B^i B^j 
+ \frac{1}{2k} \left((F^4)_{asd}\right)^2,
\ee
which, modulo a surface term, is just the Plebanski action for General Relativity plus the action of Maxwell theory. 

The reality conditions that need to be imposed to get a Lorentzian signature theory are unchanged in the $SU(2)$ sector, and are given by (\ref{reality}). The additional reality condition that needs to be imposed is that the $U(1)$ connection is real. This can be done by requiring
\be\label{reality-B4}
B^4 \wedge (B^4)^* =0.
\ee
Indeed, we know from (\ref{B4-F4}) that on-shell $B^4$ will be purely anti-self-dual. Then the condition (\ref{reality-B4}) says that this anti-self-dual 2-form is the anti-self-dual part of a real 2-form, which then requires $A^4$ to be real. So, the reality condition for the $U(1)$ sector takes a form similar to the conditions (\ref{reality}) in the gravity sector, which is nice. 

All in all, the theory (\ref{action-Robinson}) is probably the nicest known way of putting together GR and Maxwell theory. It does so just by enlarging the structure already present in the formulation of pure GR, and the constructions used are quite analogous to what is present in the case of pure GR. We cannot think of any drawback of this unification scenario, except that it does not generalise in any natural way to YM theory, as we review next. 

\subsubsection{Generalisation to Einstein-Yang-Mills}

The action (\ref{action-Rob-comp}) can be trivially generalised to give gravity plus YM. To this end, one just needs to introduce extra indices. Let $a,b,\ldots$ be indices for the Yang-Mills gauge group. We can then write
\be\label{action-Rob-YM}
S[A,B,\Psi,\phi] = \int B^i F^i  + B^a F^a - \frac{1}{2} \left( \Psi^{ij} + \frac{\Lambda}{3} \delta^{ij}\right)B^i B^j   - \frac{1}{2} k B^a\wedge B^a - \phi^{ia} B^i B^a.
\ee
Exactly the same procedure of integrating out $\phi^{ia}$ and $B^a$ gives that $B^a$ is anti-self-dual and a multiple of the anti-self-dual part of $F^a$. This results in the action of Einstein-Yang-Mills theory in the form
\be
S[A^i,B^i,\Psi, A^a] = \int B^i F^i - \frac{1}{2} \left( \Psi^{ij} + \frac{\Lambda}{3} \delta^{ij}\right)B^i B^j 
+ \frac{1}{2k} \left((F^a)_{asd}\right)^2.
\ee
Now $k$ receives the interpretation of a multiple of the YM coupling constant. 

On the other hand, the action (\ref{action-Rob-YM}) can be written in the form similar to (\ref{action-Robinson})
\be\label{action-Rob-YM-1}
S[A,B,M] = \int B^I F^I - \frac{1}{2} M^{IJ} B^I B^J,
\ee
where the index $I=(i,a)$ and the matrix $M^{IJ}$ is required to be of the form
\be\label{M-form}
M^{IJ} = \left( \begin{array}{cc} \Psi^{ij} + \frac{\Lambda}{3} \delta^{ij} & \phi^{ia} \\ \phi^{ia} &  k\delta^{ab} \end{array} \right).
\ee
Unfortunately, this way of writing the action shows that we are not really dealing with a unified theory. First, the gauge group in the above is just the product of the gauge groups $SO(3)\sim SU(2)$ required to get GR in Plebanski formalism and the YM gauge group. Second, it is very hard to motivate the form of the matrix $M^{IJ}$, as (\ref{M-form}) requires that only the trace part be present in the lower-diagonal block of this matrix. This form can be imposed with the help of $1+n(n+1)/2$ Lagrange multipliers, where $n$ is the dimension of the YM gauge group, but this is unattractive. So, overall we must conclude that (\ref{action-Rob-YM-1}) with (\ref{M-form}), while giving a way of rewriting the Einstein-Yang-Mills system Lagrangian, is not really a unification scheme. 

\subsubsection{More general unified models}

The models studied in \cite{TorresGomez:2009gs}, \cite{TorresGomez:2010cd} and \cite{Krasnov:2011hi} can all be described from a viewpoint similar to the one previously discussed. Thus, we shall consider the action of the same general type (\ref{action-Rob-YM-1}), but add to it a single constraint on the matrix $M^{IJ}$
\be\label{action-chiral-unified}
S[A,B,M] = \int B^I F^I - \frac{1}{2} M^{IJ} B^I B^J + \mu ( f(M) - \lambda).
\ee
Here the gauge group is arbitrary, and can be taken to be simple, and $I,J,\ldots$ is the Lie algebra index. The matrix $M^{IJ}$ has values in the second symmetric power of the Lie algebra. The model is specified by choosing the function $f(M)$, which is assumed to be a gauge-invariant. If desired, one can impose on the matrix $M^{IJ}$ more than one constraint, as in (\ref{action-Robinson}), but in all models \cite{TorresGomez:2009gs}, \cite{TorresGomez:2010cd} and \cite{Krasnov:2011hi} just a single constraint was imposed. 

The main idea of the analysis in the papers \cite{TorresGomez:2009gs}, \cite{Krasnov:2011hi} was to choose an appropriate background that breaks the symmetry to the gravitational $SU(2)$ times the subgroup that commutes with this $SU(2)$. Such a background can be specified by choosing an embedding of $SU(2)$ into the full gauge group. One can then expand the action (\ref{action-chiral-unified}) around the background chosen, and see what are the propagating modes. In this analysis one does not need to make any assumptions on $f(M)$ apart from some generality. It is found that the $SU(2)$ sector describes gravitons, the sector charged under the subgroup that commutes with the gravitational $SU(2)$ describes massless gauge fields, and what can be called off-diagonal sector describes exotic fields that are charged under Lorentz as well as the YM group. The main difficulty of the models of this type is that at the non-linear level what arises is a modified gravity of the type considered in \cite{Krasnov:2006du}, and unlike the case of the Plebanski formalism for GR, the reality conditions to be imposed on the fields to get Lorentzian signature metrics and real Lagrangian are not under control.

\subsection{Exceptional unification}

A bold attempt at unification that attracted much public attention
was the ``exceptionally simple theory of everything'' 
based on the group $E_8$ \cite{Lisi:2007gv}.
The original idea was to fit all known particles
into the 248-dimensional adjoint representation of $E_8$.
The proposal met with skepticism, which finally
crystallized in a paper giving some no-go theorems
\cite{Distler:2009jt}.
Oversimplifying, the theorems can be summarized as:
\begin{enumerate}
\item there cannot be three fermion families in the adjoint of $E_8$;
\item there cannot be one chiral fermion family in the adjoint of $E_8$.
\end{enumerate}
Let us spell them out in some more detail.
 
The first theorem can be proven as follows. 
Fermions are spinors, and therefore change sign under a 360 degree rotation.
When one embeds the spin group $SL_\C(2)$ in $E_8$,
the rotation by 360 degrees corresponds to a central element whose square is 1.
In a spinor representation, such an element must act as minus the identity.
One can use results of Cartan to the effect that the subspace of the Lie algebra of $E_8$
where this element acts as minus the identity has dimension 112 or 128, depending on the real form.
Since one spinor family or antifamily (including a right handed neutrino) has real dimension 64,
there can be at most two families/antifamilies in the adjoint of 
$E_8$.
 
The second and stronger result of Distler and Garibaldi is based on chirality.
For a given embedding of $SL_\C(2)$ in (a real form of) $E_8$, 
define the GUT group to be the centralizer of $SL_\C(2)$.
If the fermions happened to be in a real or pseudoreal representation of this group, they could not be chiral.
Distler and Garibaldi worked out the complete list of all GUT groups that could be embedded in
(real forms of) $E_8$, and of the corresponding fermionic representations: they are all real or pseudo-real.
As a consequence, $E_8$ unification as proposed by Lisi 
predicts a nonchiral fermion spectrum.
Statement (2) above then follows if one makes the assumption that all the fermions of one family must be
in a chiral representation of the GUT group, as the known particle spectrum demands.
Unfortunately (or fortunately, depending on one's taste) there is 
some wiggle room here: one cannot exclude
with absolute certainty the existence of additional families or antifamilies with large masses.
Thus, one could take one known family and an antifamily, corresponding to one complex representation and its conjugate,
and together they would form a real representation, which may happen to occur in the Distler-Garibaldi list.

In fact, in \cite{Lisi:2010uw} Lisi describes an embedding
of the ``graviGUT'' group $SO(3,11)$ of \cite{gravigut} in $E_8$,
which could be extended to an embedding of a fermionic family 
and the corresponding antifamily
(see also \cite{Douglas:2013gua} for a more precise
description of the algebras involved).
One would then have to find a mechanism that gives very large masses to all the particles in the antifamily,
while those in the family remain light. 
 
This is hard, but in our opinion the central issue is another one. 
Even if there was a physically believable mechanism
to get rid of the antifamily without contradicting known experimental facts, how would we account
for the presence of three families in nature?
Given the first result of Distler and Garibaldi, there are only two possibilities. The first is to give up the constraint
that all particles must be contained in a single copy of the adjoint of $E_8$. For example, one could take
three adjoints - but then one would also have three copies of the electromagnetic field, three copies of the
gravitational field and so on, and we certainly don't want this.
Or, one could put the fermions in a larger representation. But since the gauge fields must be in the adjoint, this
means that one would have fermions and bosons in different representations. This is normal in GUTs, but is contrary
to the spirit of Lisi's original idea, and furthermore the profusion of unwanted particles would become even bigger.

The second possibility is to try to evade Distler and Garibaldi's first no-go theorem by changing the rules of the game.
This is essentially what Lisi tries to do in
\cite{Lisi:2007gv,Lisi:2015oja},
where he suggests that the three families could
be related by triality.
Three 64-dimensional subspaces in the Lie algebra of $E_8$,
related by automorphisms, 
would each be identified as a spinor representation 
of a {\it different} $SL_\C(2)$ subgroup of $E_8$.
This departs from the framework of unified theories
that we spelled out in section II.A,
where the order parameter selects the unbroken subgroup, and all particles fall in specific representations of this 
{\it fixed} subgroup.

\section{Discussion}
\label{sec:disc}

We begin by summarizing, in subsection A, the strenghts and weaknesses
of the unified models, in particular those based on the 
McDowell-Mansouri and BF formulation.
Subsection B contains a discussion of possible relations
between KK theories and the unified theories based on extensions
of the internal space.
The following three subsections contain some comments on other aspects
of the theory that we had not touched upon previously:
the role of the Coleman-Mandula theorem,
quantization and the nature of the unified phase.
Subsection F contains our conclusions.

\subsection{Discussion of the unified models}

A general feature that is shared by all the unification schemes of ``enlarging the gauge group'' type
is the appearance of fields that transform with respect 
to both Lorentz and YM gauge groups.
These are akin to the  leptoquarks of the GUTs,
and their appearance seem unavoidable in any model where the 
Lorentz and YM groups are embedded into a larger gauge group. 
Indeed, it is intuitively clear that the ``off-diagonal'' 
components of the fields must transform under both, 
and so one will obtain exotic fields of a type not yet seen in Nature.
In some of the models it is clear that these fields can be made very massive. In others, a detailed understanding 
of their fate is still lacking.

We can divide the unified models of ``enlarging the gauge group'' type into two categories.
On one hand we have the Einstein-Cartan-type unified
theories discussed in section \ref{sec:ecu_bose}.
As we have seen, it is not possible to preserve the
polynomial character of the Einstein-Cartan theory in the
extended, unified, models. In particular, these models cannot be written in terms of differential forms.
Modulo certain possibilities to be discussed in subsection \ref{sec:quantum} below,
they should be viewed as effective field theories valid below
the Planck scale, much like GR itself.
This is somewhat disappointing, because it means that they
can only describe the ``broken'', or ``Higgs'' phase of
the theory. They do not provide a description of the ``unified'' phase
and therefore do not fulfil all the requirements
that we spelled out in section \ref{sec:PP-unific}.
On the other hand, they clearly indicate the nature of
the order parameter and also give a satisfactory description
of the fermionic sector.
It is not at all obvious that such a description would
have been possible. 
In particular, the fact that the fermions that exist in nature
form the simplest representations of the unified ``graviGUT'' 
group $SO(3,11)$, and the fact that one can write an action
for them that reduces to the correct one in the broken phase,
are among the strongest indications that there may be some truth
in this approach to unification.

If one insists for a polynomial decription of the bosonic variables
at the fundamental level,
then one has to turn to the second type of models,
those based on the MacDowell-Mansouri or BF
formulations, to which the rest of this section is devoted.

Let us first remark that once larger gauge groups than required to get gravity are considered, there is little difference between the non-chiral unified model (\ref{action-FdP-mod}) and the chiral one (\ref{action-chiral-unified}). Indeed, one should just interpret the index pair $IJ$ in (\ref{action-FdP-mod}) as the Lie algebra index $I$ in  (\ref{action-chiral-unified}). Then the fact that the matrix that appears in front of $B^{IJ} \wedge B^{KL}$ in (\ref{action-FdP-mod}) is of the type specified can be imposed as a constraint of the type present in (\ref{action-chiral-unified}). 

The only difference that arises between these two types of unification schemes is in the natural background to expand about. In the non-chiral models the natural background is taken to be (\ref{B-vac}). In this background it is the subgroup $SO(4)$ of the full gauge group that gets interpreted as the gravitational one, and what commutes with it as the YM gauge group. In the chiral models it is more natural to take a background in which only one of the two chiral halves of $SO(4)$ is ``switched on''. On such backgrounds only an $SU(2)$ subgroup of the full group describes gravity, while what commutes with it describes YM. A proposal along these lines has been also made in \cite{Alexander:2012ge}, where the second $SU(2)$ inside $SO(4)$ was proposed to describe the weak gauge group. However, as we have already mentioned, in chiral interpretations of the theory (\ref{action-chiral-unified}) one wants to allow all fields to be complex, with some suitable reality conditions imposed to select a sector with Lorentzian metrics and real action. Unfortunately, such reality conditions are in general not understood, and so the chiral interpretation of the model (\ref{action-chiral-unified}) suffers from this ambiguity in how to select the appropriate ``real slice''. On the other hand, as we already discussed, if the model (\ref{action-chiral-unified}) is to be interpreted as a non-chiral one, with all fields real, then at least its $SO(4)$ sector is likely to have propagating degrees of freedom with wrong sign kinetic terms. There may be ways out of this by imposing more than one constraint on $M^{IJ}$, but this has not been studied. 

Having pointed out that there is no substantial difference between the unified models of BF-type, we can list some general features that are shared by the models of MacDowell-Mansouri and BF-type. First, in all these models, after the gauge group is enlarged, it is no more clear what is the spacetime metric. We have seen that in the MacDowell-Mansouri case there are several different frame fields after the gauge group is enlarged. It is no more clear how the ``physical'' metric is constructed from them. In a similar fashion, in the models of BF-type it is only after a background is selected and the theory is expanded around it that one can point out the variables that describe gravitons. In the full non-linear regime it is impossible to select which combination of fields plays the role of ``the metric'' in these theories. This is not necessarily a drawback of these unification schemes, as it may be a true feature of the unification, but it should be kept in mind.

Another property that was already mentioned is that in all these scenarios nothing forces us to restrict our attention to orthogonal groups. One can of course make this restriction, having the desired fermion transformation properties in mind. If one had hopes for a unique ``theory of everything''
this may come as a disappointment.
On the other hand, this is not worse than ordinary YM theories,
where one is free to choose the gauge group to fit the
observed particle multiplets.

Finally, even if an interesting bosonic model is constructed by following one of these unification schemes, one will still face the question of how to couple fermions to it. The models of MacDowell-Mansouri type face less problems in this regard, because some components of the connection receive the interpretation of the frame field that the fermions can couple to. In contrast, in models of BF-type there is no more a frame field. The metric-like variable is now a Lie algebra valued two-form. It is not easy to couple fermions to two-forms, with the only known result in this direction being described in \cite{Capovilla:1991qb}. But the coupling mechanism of this reference is only known to work for the case of Plebanski, and is unlikely to work for the generalised models in which the two-form field no longer satisfies the $B^i\wedge B^j \sim \delta^{ij}$ simplicity constraint. So, at least at present, the coupling to fermions appears problematic for the BF-type models. 

Our final remark is that in all these scenarios the symmetry breaking would be caused spontaneously, by selecting a particular solution of the field equations. As we already emphasised, there can be different symmetry breaking patterns depending on how the ``gravitational'' gauge group gets embedded into the full gauge group by the background solution. So, this would mean that different phases of the theory appear as different solutions of the dynamical equations. 
Unlike the usual particle physics Higgs mechanism
(see point 4.a in section \ref{sec:whatis}), there seems to be no potential to select one as being energetically favoured over another.

\subsection{Extending the gauge group and Kaluza-Klein}
\label{sec:eggkk}

We have presented the unified theories that enlarge the gauge group as sharply different from Kaluza-Klein theory. Indeed, one may say that they are ideologically opposite: in Kaluza-Klein theory spacetime structures (components of the metric) are dynamically converted into internal structures
(gauge and Higgs fields). In extending the gauge group scheme, internal structures
(fiber metric and connection) are dynamically converted into spacetime structures
(spacetime metric and connection). In the former, the focus is always on the metric
and in this they fit the Einstein's view of gravity, 
whereas in the latter the focus is more on the connection, 
treated as an independent variable, a point of view 
that is closer to Cartan's.

In spite of this difference, one can think of several ways of
relating the two approaches. The simplest such relation comes from performing the Kaluza-Klein dimensional reduction while using the higher-dimensional vielbein formalism. In this formalism the basic fields will be a co-frame, which is a one-form on some higher-dimensional manifold $M^n$ with values in $\R^{(p,q)}$ for some values of $p,q: p+q=n$. There is also the spin connection, which is locally a one-form on $M^n$ with values in the Lie algebra ${\mathfrak so}(p,q)$. When we dimensionally reduce to four dimensions, some of the components of these fields become scalars from the 4D point of view, while the other components give rise to 4D one-forms with values in either $\R^{(p,q)}$ or the Lie algebra ${\mathfrak so}(p,q)$. But one will also obtain such 4D one-forms with values in a big Lie algebra by starting with the 4D Einstein-Cartan formulation and ``enlarging'' the gauge group. This shows that at the level of kinematics the Kaluza-Klein higher dimensional theory (in the frame formalism) contains the fields of the ``enlarging the gauge group'' approach as a subset. At the same time, at the dynamical level this relation disappears: In the Kaluza-Klein context the connection is completely determined by the frame and its derivatives, which is not the case in the ``enlarging the gauge group'' approach. 

Another relation between the two approaches can be traced to papers of Weinberg who discussed a generalization of Kaluza-Klein theory where the higher dimensional gravity that one starts with
is not ordinary gravity (described by a metric)
but rather a theory that has a different invariance group
$G\subset GL(4+N)$ \cite{Weinberg:1984dq,Weinberg:1984ke}.
In the language of modern differential geometry,
one would say that the higher-dimensional 
tangent bundle has a $G$-structure.
Ordinary gravity corresponds to the case $G=SO(p,q)$
with $p+q=4+N$.
One then assumes spontaneous compactification to Minkowski
times a manifold with an isometry group $C_M$ acting
transitively on the $N$-dimensional space.
The requirement that $G$ contains the Lorentz subgroup
$O(3,1)$ leads to $G$ being a direct product of a 
higher-dimensional Lorentz group times a group $G'$.
Among all possible choices there is also the case
$G=SO(3,1)\times SO(10)$.

For instance, the appearance of this specific $G$-structure in a
higher-dimensional theory can be achieved by coupling the connection
to bosons with values in the Grassmannian
$\frac{SO(3,11)}{SO(3,1)\times SO(10)}$
of 4-dimensional planes in 14 dimensions (with fixed signature
for the induced metric). Explicitly, the Grassmannian field
can be described
by giving 4 linearly independent vectors $e^a_\mu$, $a=1,2,3,4$, and $\mu=1,\ldots,14$,
modulo Lorentz transformations.
As a quick check,
the dimension of this space is 56 (the number of components
of $e^a_\mu$) minus 10 (the number of orthonormality constraints) minus 6 (the dimension of the Lorentz group).
This is indeed equal to the dimension of the coset space.

Now let the $14$-dimensional Christoffel symbols $\Gamma_\rho{}^\mu{}_\nu$, spin connection $\omega_\mu{}^I{}_J$
and frame field $\theta^I{}_\mu$ be related as in (\ref{spinconn}).
Then, imposing the condition
\be
D_\mu e^a{}_\nu=0\ ,
\label{giulia}
\ee
where $D$ is the total covariant derivative,
reduces the gauge group to the desired $G$.
To see this, note that the matrix $e^a{}_b=e^a{}_\mu\theta_a{}^\mu$
has rank four and acts as projector in the subspace
spanned by the four vectors $e^a$.
Due to (\ref{spinconn}), 
\be
D_\mu e^a{}_b=0\ .
\label{giulia2}
\ee
This is equivalent to (\ref{giulia}).
We can choose the frame field such that the first four vectors
coincide with the vectors $e^a$.
In this gauge $e^a{}_b=\delta^a_b$ for $a,b\in(1,2,3,4)$
and zero otherwise.
Then equation (\ref{giulia2}) implies that the mixed components
$\omega_{\mu IJ}$ vanish, while those in the diagonal
$4\times 4$ and $(N-4)\times(N-4)$ blocks remain free.

Alternatively, one can treat these vector fields as dynamical
and add to the action a kinetic term that is square
in $D_\mu e^a{}_\nu$.
The condition (\ref{giulia}) then appears as a property
holding at low energy.
Either way, upon dimensional reduction to four dimensions
this would
lead to a model containing the same low energy fields
as the ones of our Einstein-Cartan-type unified model
of sections IV.A.2-3-4.

Another possible relation goes via brane-world scenarios
\cite{Faddeev:2009dy,Faddeev:2009gm}.
In this case four-dimensional spacetime would be embedded
in a $4+N$-dimensional space with target space
coordinates $Y^I(x)$, where $x$ is a coordinate in four dimensions. 
The induced spacetime metric would be given by $g_{\mu\nu} = \eta^{IJ} \theta_{I\mu} \theta_{J\nu}$ where $\eta^{IJ}$ is the metric in the target space and 
$$
\theta^I{}_\mu=\partial_\mu Y^I\ .
$$
This is a ``generalised'' tetrad of the type we considered in sections IV.A.2-3-4. The peculiar feature of this scenario is that the soldering form satisfies the condition $\partial_\mu\theta^I{}_\nu-\partial_\nu\theta^I{}_\mu=0$.

\subsection{The Coleman-Mandula theorem}
\label{sec:colemandula}

It has sometimes been said
that the Lorentz group cannot be unified with
a YM gauge group, due to the Coleman-Mandula
theorem \cite{Coleman:1967ad}, whose folk version states that 
``one cannot mix internal and spacetime symmetries''.
Of course, this is too broad a statement.
The theorem itself has several hypotheses,
the first and most relevant one being Poincar\'e invariance.

The unified theories that enlarge the gauge group,
in their unified phase (to be determined),
would violate even a much weaker version of this hypothesis,
namely the existence of a nondegenerate metric.
For it is only when the VEV of the soldering form vanishes
that the gravitational and non-gravitational interactions
would be truly unified.

In the broken phase, assuming that the VEV of the metric is 
flat Minkowski spacetime,
the global symmetry group of the theory would indeed be 
the product of the Poincar\'e group and $O(N)$,
as required by the theorem.

\subsection{Quantization}
\label{sec:quantum}

So far, we have concentrated on the classical aspects
of unified theories. If we tried to turn these into quantum theories,
we would face the same problems that are encountered for pure gravity. 
Namely, the models where one enlarges the gauge group,
while spacetime remains four-dimensional,
have the same types of divergences that are encountered
in gravity and are therefore power-counting non-renormalizable.
In the case of higher-dimensional Kaluza-Klein theories the divergences are even worse.
So, all these unified theories can be assumed to be UV incomplete. 

There are several possible attitudes towards this issue.
A modest attitude would be to view them as effective field theories, valid up to some energy scale. 
However, unification typically becomes manifest only at the Planck scale and this is precisely the scale where effective theories of gravity break down. Thus, this point of view seems to confine the unified theories
to the domain where they are least interesting.
This seems indeed to be the case for the
non-polynomial formulations of section \ref{sec:ec}.

However, given that a unified theory of this type would describe all known forces of Nature, one clearly wants to set the bar higher.
Then, aside from string theory, 
there are essentially two possibilities. 
One is provided by asymptotic safety \cite{weinberg,reuterrev,perbook2}. 
The other one is to hope for some kind of miracle (or more appropriately, for some yet to be identified symmetry principle) that would make a very special theory of this sort UV complete. For example, there is a hope that $N=8$ supergravity in four dimensions may be quantum finite, in part due to its very high degree of supersymmetry, in part due to mysterious enhanced cancellations, see e.g. \cite{Bern:2014sna}. So, it may be that there exist very symmetric power-counting non-renormalisable theories that still manage to make sense quantum mechanically. None of the described above schemes qualifies as a ``very symmetric'' theory, but it is not excluded that there are better classical unification scenarios yet to be discovered. Thus, in our opinion, the power-counting non-renormalisability of all the scenarios that have been considered so far should not prevent researchers from looking for more elegant classical unification schemes.

\subsection{The unified phase}
\label{sec:unified}

A unified theory should be able to describe
both a ``unified'' or ``high energy'' phase and
a ``broken'' or ``low energy'' phase. Only in the low energy phase it would be possible to distinguish which fields are gravitational and which represent ``matter''. No such distinction would be possible in the unified phase, with all known fields (and possibly some new ones) being components of (most optimistically) a single field. 

As we have seen, in all the schemes that enlarge the gauge group it is
the soldering form (or a field that plays similar role) that 
acts as the order parameter. It is when this field assumes a non-trivial vacuum expectation value that the symmetry of the original theory is broken, and physics of the type that we see in our world arises. One can then speculate that there may be different symmetry breaking patterns, depending on the VEV that the metric-like field assumes. Further, one can imagine such different symmetry breaking patterns being realised dynamically, e.g. by a process in which the theory moves from one possible vacuum solution into another. For example, scenarios of this type are possible in the context of models of BF-type, see \cite{Krasnov:2011hi}. 

The above leads to the speculation that the unified phase is one where the vacuum expectation value of the metric field (or field that plays similar role) is zero. It thus appears that the unified phase should be described by a ``topological theory'', whatever that may mean in this context. 
Indeed, some formulations of GR are strongly suggestive of this. For instance, in BF-type formulations, the action of the theory always is the sum of the kinetic BF term, which, taken by itself would give a topological theory, and a ``potential'' term for the B-field that breaks the topological symmetry. It is tempting to speculate that this is the topological BF term that describes the ``unified'' phase. However, it's difficult to see what kind of calculation could support such a speculation. 

\subsection{Conclusions}

We now know that all interactions except gravity are
correctly described by YM theories.
Aside from the choice of gauge group, these are the unique low-energy theories of spin-one fields.
Some of these theories, such as QCD, are UV complete.
Others, such as the abelian sector of the SM,
are not UV complete because they lack an UV fixed point.
The simplest option is to assume that they are embedded in
a non-abelian, asymptotically free grand-unified gauge theory.

On the other hand we have gravity, which is correctly described by GR, which is the unique low-energy theory of spin-two fields.
In its most familiar formulations (in terms of metric or vierbein)
it is a nonpolynomial theory showing striking similarities
to the gauged non-linear sigma models. 
Insofar as these models are the universal low-energy description
of some symmetry-breaking phenomenon,
this suggests that gravity, as we know it,
is also the relic of some symmetry breaking phenomenon occurring
at the Planck scale.
But there are also other polynomial formulations of the theory
that, in a way that we presently don't understand,
may provide a more fundamental description of gravity.

These are the two theories that we try to unify.
We have seen in section \ref{sec:ec} that, if we content ourselves with the low-energy description,
such a unification is possible. It is essentially GR coupled to $SO(10)$ YM fields and fermions,
written in an $SO(3,11)$-invariant way.
While suggestive, this is not fully satisfactory,
because the unified theory is supposed to describe physics
also above the unification scale. We have also described attempts to go beyond this effective description, based on the use of the MacDowell-Mansouri and BF formulations. 

While both GR and YM are unique low energy theories, probably the most serious drawback of all the unification attempts that we have considered is that the uniqueness is lost: there are ambiguities in how to write the Lagrangian, and typically many terms compatible with all the symmetries can be written, with many new coupling constants.\footnote{A notable exception is the Kaluza-Klein scheme based on 11-dimensional supergravity, which is a theory with very strong uniqueness properties. But this scheme has other difficulties, as previously discussed.} 
This is probably a sign that none of these theories,
taken by itself, should be taken too seriously.
At the same time, the partial successes of these attempts, 
taken together, suggest that the classical unification of gravity 
with YM and other known fields may be possible.

\medskip

\subsection*{Acknowledgements} KK would like to thank SISSA for hospitality during a visit when this project has started. At the early stages of work on this project KK was supported by ERC Starting Grant 277570-DIGT\@.

\end{document}